\RequirePackage{fix-cm}
\documentclass[twocolumn,epjc3]{svjour3}  
\smartqed  
\RequirePackage{graphicx}
\usepackage{amsmath}
\usepackage{amssymb}
\usepackage{slashed}
\usepackage{cancel}
\usepackage{cite}
\usepackage{float}
\usepackage[symbol]{footmisc}
\usepackage[singlelinecheck=false,justification=centering]{caption}
\usepackage{physics}
\usepackage{appendix}
\usepackage{xcolor}

\newcommand{\ba}{\begin{aligned}}
\newcommand{\ea}{\end{aligned}}

\DeclareMathAlphabet{\mathpzc}{OT1}{pzc}{m}{it}
 \csname
@addtoreset\endcsname{equation}{section}

\journalname{Eur. Phys. J. C}
\begin{document}

\title{Axions in a highly protected gauge symmetry model}

\author{Q. Bonnefoy\thanksref{e1,addr1}, E. Dudas\thanksref{e2,addr1}
        \and
        S. Pokorski\thanksref{e3,addr2} 
}

\thankstext{e1}{e-mail: quentin.bonnefoy@polytechnique.edu}
\thankstext{e2}{e-mail: emilian.dudas@polytechnique.edu}
\thankstext{e3}{e-mail: stefan.pokorski@fuw.edu.pl}


\institute{Centre de Physique Th\'eorique, \'Ecole Polytechnique, CNRS, Universit\'e Paris-Saclay,\\ Route de Saclay, 91128 Palaiseau, France \label{addr1}
           \and
           Institute of Theoretical Physics, Faculty of Physics,University of Warsaw,\\ ul. Pasteura 5, PL-02-093 Warsaw, Poland \label{addr2}
}

\date{Received: date / Accepted: date}

\maketitle

\begin{abstract}
We study QCD axion or cosmological axion-like particles (ALPs) in a model inspired by the recent interest in 4-dimensional clockwork models, with the global symmetry being accidentally enforced by a gauge abelian quiver with scalar bifundamental fields.

For the QCD axion, we analyze the connection between the degree of protection of the axion mass against gravitational corrections, the explanation of the hierarchy $f_a \ll M_P$ and the number of colored fermions needed to generate anomalous couplings to gluons, all linked together by the underlying gauge symmetries. Based on that model and on the comparison with earlier models in the literature, we derive certain general conclusions on QCD axion models that use accidental global symmetries.

For the ALPs, assuming that their mass is solely given by gravitational corrections, we identify the parameter space where the decay constant and the mass are consistent with the DM abundance, and we show that this clockwork-inspired model is a particularly economical model for a very light ALP DM candidate.
\end{abstract}

\section{Introduction}

New pseudo-Goldstone bosons (PGB's) may play an important role in particle physics and cosmology, since they can solve the strong CP-problem (QCD axion) \cite{Peccei:1977hh,Weinberg:1977ma,Wilczek:1977pj} and/or explain dark matter \cite{Preskill:1982cy,Abbott:1982af,Dine:1982ah}, drive inflation \cite{Freese:1990rb,Kim:2004rp} or make dark energy dynamical \cite{Ratra:1987rm,PhysRevLett.75.2077,Kim:2002tq,Kim:2013pja}. The PGB playing the role of the QCD axion must have anomalous couplings to gluons whereas such couplings are not needed for the axion-like particles (ALPs) that only play the latter roles. However, in both cases one is facing several, partly similar, issues.

One is that the PGB's must be generically very light so there is a need to protect the global symmetries from a too large explicit breaking by gravitational corrections\footnote{There can be other sources of such a breaking but we focus on gravitational corrections.} \cite{Hawking:1987mz,Giddings:1988cx,Banks:2010zn}. If an axion is to solve the strong CP-problem, the non-anomalous explicit breaking must be subleading and its mass is approximately determined by the confinement scale and the axion decay constant $f_a$. For an ALP, the most economical possibility is that its mass is just given by the gravitational corrections, the assumption we make in this paper. The question about the proper protection of the axion and/or ALP global symmetries has been addressed by many authors \cite{Kim:1981bb,Georgi:1981pu,Dimopoulos:1982my,Kang:1982bx,PhysRevD.46.539,Kamionkowski:1992mf,Holman:1992us,Hill:2002me,Hill:2002kq,Dias:2002gg,Harigaya:2013vja,Dias:2014osa,Ringwald:2015dsf,Redi:2016esr,Fukuda:2017ylt,Lillard:2017cwx}. In the field theoretical models in four dimensions, one often considers the symmetry from which the PGB's originate as an accidental consequence of gauge symmetries, i.e. as unbroken by any gauge-invariant operator up to a given dimension. Typically, strong enough protection requires  either large charges of the scalars under the gauge symmetry(ies) (see e.g. \cite{PhysRevD.46.539}) or many gauge groups as in quiver models (see e.g. \cite{Hill:2002me,Hill:2002kq}). The latter can be viewed as inspired by the latticized versions of extra-dimensional models where the  PGB's  can be interpreted as fifth components of vector fields which appear as scalars in 4d.

Another issue  is the origin  of the scale $f_a$  and of the potential hierarchy $f_a\ll M_P$. Such a hierarchy is required for the generic  QCD axion window  but not needed for the ALPs as dark matter (DM) candidates only, or even not acceptable for $m_\text{ALPs}\sim O(10^{-15}-10^{-20})$ eV. One more difference between  the QCD axion and the ALPs models is that the former requires a set of colored fermions to generate the anomalous couplings to gluons. Thus, the constraints are different for the two cases.

In explicit models for a protected QCD axion, one can connect the degree of protection of the axion mass, the explanation of the hierarchy $f_a\ll M_P$\footnote{In most models with good protection the scale $f_a$ is not identical to the original scale $f$ of the spontaneous global symmetry breaking and one may consider the possibility of $f_a\ll f\sim M_P$.} and the number of colored fermions and their masses, all linked together by the underlying gauge symmetries. In DM ALPs models, assuming that their mass is solely given by gravitational corrections, one can identify the parameter space such that the scale $f_a$ and the mass $m_a$ combine to give the observed relic abundance.

In this paper we discuss  those questions using as our laboratory a model inspired by the recent interest in 4-dimensional clockwork models \cite{Choi:2015fiu,Kaplan:2015fuy,Giudice:2016yja,Ahmed:2016viu,Hambye:2016qkf,Craig:2017cda,Coy:2017yex,Giudice:2017suc,Choi:2017ncj,Giudice:2017fmj,Teresi:2018eai}. Based on that model and on the comparison  with earlier models in the literature, we derive certain general conclusions on the QCD axion models that use global symmetries that are consequences of gauge symmetries. Secondly, we show that the clockwork inspired model is a particularly economical model for a very light ALP DM candidate.

Our model is the 4d quiver model obtained by latticizing a 5d (abelian) gauge theory in a linear dilaton background \cite{Antoniadis:2001sw}, with Dirichlet boundary conditions for the 4-d components of the gauge boson \cite{Ahmed:2016viu,Coy:2017yex}. As a result of the 5d gauge invariance, the 4d field content is such that its most general renormalizable gauge-invariant lagrangian preserves an accidental global symmetry. Furthermore, the specific 5d background, or equivalently the specific 4d gauge charge assignment, ensures a strong protection of this accidental symmetry from explicit breaking terms, even when the discretization is crude (i.e. when the quiver has few sites), and it generates a hierarchy between the effective axion decay constant $f_a$ and the scale $f$ of spontaneous symmetry breaking: $f_a$ is reduced by a factor which grows exponentially with the number of quiver sites, in a way opposite to the usual clockwork models. This effective scale $f_a$, which appears in the potential of the axion and its couplings to gauge fields, is not the only scale parametrically different from $f$: the axion has a clockwork profile along the quiver sites, and this profile can generate effective coupling scales which are bigger than $f$ when one considers for instance couplings to the spins of matter particles.

Before we proceed, let us recall that there are many arguments, from black hole ones to string theory ones, suggesting that global symmetries are broken by Planck scale effects. The strength of the breaking is well defined in a consistent theory of quantum gravity. In this paper, we will parametrize gravitational corrections as higher dimensional operators in the effective theory, suppressed by powers of the Planck scale with order one coefficients, assuming that the breaking is described correctly by the EFT approach. One may wonder whether such contributions could come from non-perturbative effects and consequently enjoy a greater suppression, as suggested by studies of axions arising from antisymmetric forms in string theory \cite{Svrcek:2006yi}. However, the kind of axions discussed in this paper originate from charged matter fields. Even in string theory, those could in principle receive perturbative higher-order corrections to their potential, which would appear as usual higher-order terms in the EFT \cite{Choi:2009jt}. Furthermore, if the theory of gravity includes a heavy fermionic sector whose renormalizable couplings break the axion shift symmetry, the induced Coleman-Weinberg potential is also consistent with the effective theory point of view \cite{Hill:2002me,Hill:2002kq}. Thus, we assume in this paper that the magnitude of gravitational corrections is well described by the EFT approach, with no additional suppression.

The plan of the paper is as follows: in section \ref{4d} we recall the 4d model with a focus on the light scalar and examine its properties. In section \ref{QCD}, we discuss its potential identification with a QCD axion, analyze the interplay between the three aspects mentioned earlier, compare with other models in the literature and derive some general conclusions. In section \ref{ALPs} we consider the applications of the model to  describe cosmological light particles (e.g. PGB dark matter and quintessence models) and show that it is very economical for describing an ALP as a DM candidate. We present our conclusions in the last section. Some appendices cover additional material: appendix \ref{5d} contains the 5d deconstruction of an abelian vector field in a linear dilaton background whose low-energy limit matches that of our 4d picture, appendix \ref{massive} discusses the massive states of the model of section \ref{4d}, appendices \ref{KSVZ} and \ref{DFSZ} describe realizations of the QCD axion discussed in section \ref{QCD}, appendix \ref{calculation} displays a calculation of the axion-photons couplings of sections \ref{QCD} and \ref{ALPsDM} and appendix \ref{scanAll} discusses the ranges of parameters of the model which allow the axion to be (a detectable kind of) dark matter.

\section{Model}\label{4d}

\subsection{Gauge group and matter content}

\begin{figure*}[h]
\centering
\includegraphics[scale=0.33]{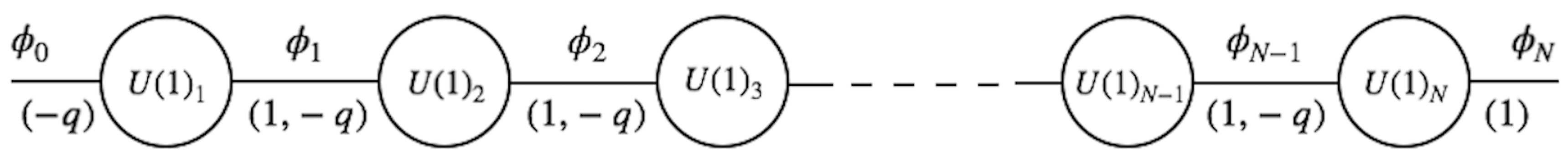}
\caption{Abelian quiver of the model}
\label{quivAxion}
\end{figure*}

The (4d) setup we consider is an abelian quiver model with bifundamental scalar fields, first presented in \cite{Ahmed:2016viu} as the deconstruction \cite{ArkaniHamed:2001ca,Hill:2000mu} of a 5d abelian gauge theory on an orbifolded linear dilaton background with Dirichlet boundary conditions for the 4d gauge field\footnote{See also Appendix A of \cite{deBlas:2006fz}.} (see appendix \ref{5d}), and whose low-energy theory was derived in \cite{Coy:2017yex} as the 4d theory obtained after chiral symmetry breaking by some confining non-abelian gauge group. The precise matter content and charge assignment is given by the quiver of Figure \ref{quivAxion} (where $q$ and $N$ are integers),
with the following (most general renormalizable) lagrangian:\footnote{We use conventions of \cite{Weinberg:1995mt}, in particular signature $(-+++)$.}
\begin{equation}
\begin{aligned}
{\cal L}=&-\frac{1}{4g_i^2}\sum_{i=1}^{N}F_{\mu\nu,i}F^{\mu\nu}_i-\sum_{k=0}^{N}(\abs{D_{\mu}\phi_k}^2+m_k^2\abs{\phi_k}^2)\\
&-\sum_{k,l=0}^{N}\lambda_{kl}\abs{\phi_k}^2\abs{\phi_l}^2 \ ,
\end{aligned}
\label{model}
\end{equation} 
where $F_i$ is the field strength of the abelian vector field $A_i$, with coupling constant $g_i$, and with the covariant derivatives $D_{\mu}\phi_k=\big(\partial_{\mu}-i(1-\delta_{k,0})A_{\mu,k}+iq(1-\delta_{k,N})A_{\mu,k+1}\big)\phi_k$. This lagrangian has a $U(1)^{N+1}$ invariance, with a $U(1)^N$ gauged subgroup.
	
This model is inspired by the so-called clockwork mechanism \cite{Choi:2015fiu,Kaplan:2015fuy,Giudice:2016yja,Ahmed:2016viu,Hambye:2016qkf,Craig:2017cda,Coy:2017yex,Giudice:2017suc,Choi:2017ncj,Giudice:2017fmj,Teresi:2018eai} and has been introduced in \cite{Ahmed:2016viu} as a possible realization of it, so we will comment on defining features of this mechanism if we recover them while we proceed, or discuss those which are different.

\subsection{Spontaneous breaking and Goldstone mode}

We are interested in obtaining Goldstone bosons, so we  consider the spontaneous breaking of the full $U(1)^{N+1}$ mentioned previously by choosing the parameters $m_k^2$ and $\lambda_{kl}$ of (\ref{model}) so that all the scalar fields $\phi_k$ get vev's $f_k$. The spectrum  then consists  after gauge fixing of $N$ massive vectors, $N+1$ massive real scalars (discussed in appendix \ref{massive}) and one Goldstone boson.

Since the vev's $f_k$ break all the gauge symmetries, $N$ out of the $N+1$ phases of the $\phi_k$ are absorbed by the gauge vectors through the Higgs effect. The absorbed phase combinations depend on the charges and vev's (we write $\phi_k=\frac{f_k+r_k}{\sqrt{2}}e^{i\frac{\theta_k}{f_k}}$):
\begin{equation}
{\cal L} \supset -A_{\mu,i}(qf_{i-1}\partial_\mu\theta_{i-1}-f_i\partial_\mu\theta_{i})\ .
\end{equation}
The last, uneaten phase $a$ remains in the spectrum after gauge fixing as a Goldstone boson associated to the accidental $U(1)_a$ global symmetry which is the ungauged factor of the $U(1)^{N+1}$ symmetry group of (\ref{model}). The profile of this boson along the original phases is orthogonal to the $qf_{i-1}\theta_{i-1}-f_i\theta_{i}$ gauge Goldstone bosons profiles. If we canonically normalize the field and the vev's are taken to be all equal, which will be assumed from now on\footnote{In the generic case, the axion profile is, up to a normalization factor, \begin{equation} a \sim \frac{\theta_0}{q^Nf_0}+\frac{\theta_1}{q^{N-1}f_1}+...+\frac{\theta_{N-1}}{qf_{N-1}}+\frac{\theta_N}{f_N} \ . \nonumber \end{equation}} (we then note $f_k=f$), it reads:
\begin{equation} 
a=\frac{\theta_0+q\theta_1+...q^N\theta_N}{\sqrt{1+q^2+...+q^{2N}}} \ .
\label{axProf}
\end{equation}
Eq. (\ref{axProf}) displays the exponential localization discussed in clockwork models, and the charges of the original scalar fields under the global symmetry also match those which appear in those models. Indeed, $U(1)_a$ acts here as $\phi_k \rightarrow e^{iq^k\alpha}\phi_k$.

\subsection{Goldstone boson protection}\label{GBprotect}

The lagrangian~(\ref{model}), has an accidental exact $U(1)_a$ global symmetry at renormalizable level, hence the axion $a$ is massless. We expect however that global symmetries are broken by gravity effects \cite{Giddings:1988cx,Hawking:1987mz,Banks:2010zn}, which forces us to include all higher order operators allowed by gauge invariance in the effective theory. For the quiver of Figure \ref{quivAxion}, these operators must be combinations of
\begin{equation}
\abs{\phi_k}^2 \text{    and    } \phi_0\phi_1^q...\phi_N^{q^N} \label{gaugeInvOp}\ .
\end{equation}
Hence, operators that explicitly break the global symmetry must involve the second term and be of extremely high dimension as soon as $q$ and $N$ are both slightly bigger than one. We thus obtain in this setup a pseudo-Goldstone boson with a mass very well protected by the gauge symmetry. The exponential dependence on $q$ and $N$ of the second operator of (\ref{gaugeInvOp}) can be used to make the boson mass ``sufficiently" small with a reasonable number of gauge groups, as we will emphasize later on. More specifically, if we use (\ref{axProf}), we find:
\begin{equation}
\begin{aligned}
&\frac{\phi_0\phi_1^q...\phi_N^{q^N}}{M_c^{1+q+...q^N-4}}+h.c.\bigg\vert_{\text{axion terms}}=\\&\hphantom{\frac{\phi_0\phi_1^q}{\phi_0\phi_1^q\phi}}2\Big(\frac{f}{\sqrt{2}M_c}\Big)^{1+q+...+q^N}M_c^4\cos(\frac{a}{f_a}) \supset -\frac{1}{2}m_a^2a^2 \ ,
\end{aligned}
\label{explicitbreaking}
\end{equation}
where
\begin{equation}
f_a = \frac{f}{\sqrt{1+q^2+...q^{2N}}}
\label{fa}
\end{equation}
and
\begin{equation}
m_a \sim \Big(\frac{f}{\sqrt{2}M_c}\Big)^{\frac{1}{2}(q+...+q^N-1)}\sqrt{1+q^2+...q^{2N}}M_c \ ,
\label{mass}
\end{equation}
and $M_c$ is the cutoff of the theory, which we take close to the Planck mass $M_P$ when we consider gravity-induced breaking effects (recall however that if  a  large number $\cal N$ of  particles is present, the actual cutoff of the theory cannot be more than roughly $\frac{M_P}{\sqrt{\cal N}}$ \cite{Veneziano:2001ah,ArkaniHamed:2005yv,Dvali:2007hz,Dvali:2007wp}). Even though $M_c$ may also be the scale of other breaking effects (such as the mass of heavy fermions explicitly breaking $U(1)_a$ and running in loops, see Appendices \ref{5d} and \ref{KSVZ} for discussions on this topic), we will for simplicity focus on gravitational scale breaking. Note that $f_a$ is significantly lower than $f$ when $N$ is large and $q>1$ (we will come back to this when discussing the QCD axion).

\section{QCD axion}\label{QCD}

We dedicate this section to the study of the compatibility of the Peccei-Quinn (PQ) idea \cite{Peccei:1977hh,Weinberg:1977ma,Wilczek:1977pj} (see also \cite{Peccei:2006as} for a review) with the setup of section \ref{4d}.\footnote{Accidental PQ symmetries have been studied in many different setups, see for example  \cite{Kim:1981bb,Georgi:1981pu,Dimopoulos:1982my,Kang:1982bx,PhysRevD.46.539,Kamionkowski:1992mf,Holman:1992us,Dias:2002gg,Harigaya:2013vja,Dias:2014osa,Ringwald:2015dsf,Redi:2016esr,Fukuda:2017ylt,Lillard:2017cwx}.}

\subsection{Accidental Peccei-Quinn symmetry in the low-energy field theory}\label{accidentalQCD}

We study in this section the low-energy effective field theory of the axion, assuming that every other massive field has been integrated out. In order to identify $U(1)_a$ with a Peccei-Quinn symmetry, we consider the following axionic coupling:\footnote{We will not pay attention to writing dimensionless quantities in the log's since it does not affect the discussion about axions which reside in the phases of the fields. Every expression in a log to appear in the rest of the paper should then be thought of as a dimensionless one (e.g. a log(scalar) means a log(scalar divided by a mass scale)).}
\begin{equation}
i\log(\phi_0\phi_1^q...\phi_N^{q^N})\Tr(G^{\mu\nu}\tilde{G}_{\mu\nu}) + h.c. \ , 
\label{gaugeInvAnom}
\end{equation}
with $G_{\mu\nu}$ the gluon field strength and $\tilde{G}_{\mu\nu}=\epsilon_{\mu\nu\rho\sigma}G^{\rho\sigma}$ its dual. In this section, we will not discuss the origin of this coupling, which may arise from a string theory or from a UV-complete field theory (section \ref{quarksInt} deals with the field theoretic case).

The operator in the log is, as we said in section \ref{GBprotect}, the first gauge-invariant term capable of coupling the axion of (\ref{axProf}) to the gluons that we could have written (using a gauge-invariant term is necessary in order not to generate any $U(1)_i\times SU(3)^2$ gauge anomaly). This coupling has two major generic features: it involves all the quiver sites, and it implies a {\it decrease} in the decay constant of the axion compared to the scale of breaking $f$. Indeed, when we plug back the axion profile (\ref{axProf}) in (\ref{gaugeInvAnom}), we obtain:
\begin{equation}
\begin{aligned}
&i\log(\phi_0\phi_1^q...\phi_N^{q^N})\Tr(G^{\mu\nu}\tilde{G}_{\mu\nu})+ h.c. \supset\\&\hphantom{i\log(\phi_0\phi_1^q)}-\frac{2\sqrt{1+q^2+...q^{2N}}}{f}a\Tr(G^{\mu\nu}\tilde{G}_{\mu\nu}) \ ,
\label{decayIncr}
\end{aligned}
\end{equation}
where we recognize the effective axion decay constant of (\ref{fa}). This suggests that the present setup could describe intermediate scale axion decay constant obtained from high scale physics (such as string scale physics).\footnote{This feature, added to the fact that each site of the quiver contributes to the anomalous coupling, is qualitatively different from those of axion clockwork models, where the anomaly is generated at one site and the effective decay constant is bigger than the scale of new physics, often considered to be $\sim$ TeV.} This feature is common to most models with global symmetry protected by gauge symmetries: the scale $f_a$ is not identical to the original scale $f$ of the spontaneous global symmetry breaking  and one may consider the  possibility of $f_a\ll f\sim M_P$. The relation between the scales $f_a$ and $f$ depends on the scalar fields charges and/or the number of gauge symmetries: specific examples are (\ref{fa}) for the model under study, (\ref{BarrSeckel}) and (\ref{HillLeibovich}) for other models present in the literature. In the model of \cite{PhysRevD.46.539}, which corresponds to $N=1$ and $q=p'/q'$, with positive coprime integers $p'$ and $q'$:
\begin{equation}
f_a=\frac{f}{\sqrt{p'^2+q'^2}} \ ,
\label{BarrSeckel}
\end{equation}
while in the model of \cite{Hill:2002me,Hill:2002kq}, which corresponds to arbitrary $N$ and $q=1$:
\begin{equation}
f_a=\frac{f}{\sqrt{N}} \ .
\label{HillLeibovich}
\end{equation}
From these expressions, we can deduce that to bring $f_a$ from the Planck scale down to the intermediate scale ($f_a \sim 10^{12}$ GeV), we require $q^N$ (respectively $\sqrt{p'^2+q'^2}$ or $\sqrt{N}$) $\sim 10^6$, for the model under study (respectively the models introduced above). This can be achieved in our model if for instance $q=3$ and $N=13$, whereas it requires $\max(p,q)\sim 10^6$ or $N\sim 10^{12}$ in the other cases discussed. An exponential hierarchy between $f$ and $f_a$ in our model hence requires a much smaller number of additional gauge groups or much smaller gauge charges than in the other models analyzed.

When non-perturbative effects of QCD turn on, (\ref{decayIncr}) induces a potential for the axion:
\begin{equation}
{\cal L} \supset m_\pi^2 f_\pi^2\cos(\frac{a}{f_a}-\theta_{\text{QCD}}) \ .
\label{QCDpotential}
\end{equation}
We also include every gauge-invariant term to the potential, according to the discussion of section \ref{GBprotect}, and in particular generate a classical explicit breaking mass term (\ref{mass}) for the axion. In order to have $\abs{\frac{a}{f_a}-\theta_{\text{QCD}}} < 10^{-10}$ at the minimum of the potential and solve the strong-CP problem, we must ensure \cite{PhysRevD.46.539,Kamionkowski:1992mf,Holman:1992us} that:
\begin{gather}
\ba
\Bigg[m_{a,\text{QCD}}& \sim \frac{m_\pi f_\pi}{f_a}\Bigg] >\\10^5& \Bigg[m_{a,\text{explicit}}\sim\Big(\frac{f}{\sqrt{2}M_c}\Big)^{\frac{1}{2}(q+...+q^N-1)}\frac{f}{f_a}M_c\Bigg] \nonumber
\ea\\
\hphantom{+++++++++}\text{or equivalently} \label{protectQCD} \\
f\lesssim \Big(10^{-5} \sqrt{2}m_\pi f_\pi (\sqrt{2}M_c)^{\frac{1}{2}(q+...+q^N-3)}\Big)^{\frac{2}{q+...+q^N+1}} \nonumber \ .
\end{gather}
For example, when $q=3,N=2$ and $M_c=M_P$, it implies $f \lesssim 10^{12}$ GeV. If now $q=3,N=3$ and $M_c=M_P$, this becomes $f \lesssim 10^{16}$ GeV.\footnote{These values show the compatibility of our setup with astrophysical ($f_a \gtrsim 10^9$ GeV) and cosmological ($f_a \lesssim 10^{11}$ GeV) bounds on the axion decay constant (the upper bound can be relaxed, if the PQ symmetry is assumed to be broken during inflation, as soon as one allows for tuning in the cosmic initial conditions for the axion), see \cite{Marsh:2015xka}.} The values of the parameters $q$ and $N$ can be of course translated into the value of the ratio $f/f_a \sim q^N$. Those numbers can be compared with those obtained in the other  models we already discussed: in the model of \cite{PhysRevD.46.539}
\begin{equation}
m_{a,\text{explicit}}\sim\Big(\frac{f}{\sqrt{2}M_c}\Big)^{\frac{1}{2}(p'+q'-2)}\frac{f}{f_a}M_c \ ,
\label{BarrSeckelbis}
\end{equation}
and for instance $f \sim 10^{12}$ GeV demands $p'+q'\gtrsim 13$. In the model of \cite{Hill:2002me,Hill:2002kq}:
\begin{equation}
m_{a,\text{explicit}}\sim\Big(\frac{f}{\sqrt{2}M_c}\Big)^{\frac{1}{2}(N-1)}\frac{f}{f_a}M_c \ .
\label{HillLeibovichbis}
\end{equation}
Then, $f \sim 10^{12}$ GeV demands $N\gtrsim 13$. Clearly, the  larger the  ratio $f/f_a$ the better the protection, but sufficient protection is obtained already for $f/f_a = {\cal O}(10)$.  In  the present set up,  this is achieved with smaller charges or smaller number of gauge groups than in the other examples described above. Unfortunately, the nice feature of the possibility of obtaining  the hierarchy $f_a\ll f\sim M_P$ in QCD axion models based on global symmetries protected by gauge symmetries is overshadowed by the fermion problem discussed in section \ref{quarksInt}.

Axion couplings to photons, which are the subject of most axion searches, are also part of this low-energy discussion. They can be derived when we consider the axionic generalizations of (\ref{gaugeInvAnom}):
\begin{equation}
\begin{aligned}
{\cal L} &\supset \frac{i}{32\pi^2}\log(\phi_0\phi_1^q...\phi_N^{q^N})({\cal C}G^{a,\mu\nu}\tilde{G}^a_{\mu\nu}+{\cal E}F^{\mu\nu}\tilde{F}_{\mu\nu})\\
&\rightarrow -\frac{\sqrt{1+q^2+...q^{2N}}}{32\pi^2f}({\cal E}-\underbrace{\frac{2{\cal C}}{3}\frac{4+m_u/m_d}{1+m_u/m_d}}_{\approx 1.92{\cal C}})aF^{\mu\nu}\tilde{F}_{\mu\nu}
\end{aligned}
\label{EandC}
\end{equation}
where $F$ is the photon field strength, $\tilde{F}$ its dual, $m_{u,d}$ are quark masses, ${\cal C}$ and ${\cal E}$ anomalous constants (which we will specify when we deal with precise models in what follows), the arrow indicates that we took into account the mixing between the axion and the mesons which arises from (\ref{decayIncr}) and (\ref{QCDpotential}) \cite{Srednicki:1985xd} and we used $m_u \approx 0.6m_d$ under the bracket in the last line. These couplings feature the dependence on the decreased effective decay constant (\ref{fa}) we already encountered in (\ref{decayIncr}). Couplings of the axion to fermions, such as axion-spin couplings, and their effective scales are discussed in section \ref{NMR}.

\subsection{Axionic couplings from heavy fermion loops}\label{quarksInt}

We now discuss the UV origin of (\ref{gaugeInvAnom}) in terms of loops of heavy fermions coupling to the axion. 

Let us first recall how axionic couplings are generated via quarks loops. (Global) anomalies with respect to $SU(3)_c$ are mediated by colored fermions\footnote{In our case, heavy fermions must obtain their mass from Yukawa couplings since Dirac or Majorana mass terms require vector-like representations.} with some charge under the (global) symmetry, which run in loops between gluons and scalars, whose phase contains part of the axion mode. The schematic procedure\footnote{An example of the triangle loop calculation, including the numerical coefficients, is presented in appendix \ref{calculation}.} is:
\begin{equation}
\begin{aligned}
&{\cal L} = -\abs{\partial \sigma}^2-\overline{Q}\gamma^\mu(\partial_\mu-iA_{\mu}^aT^a) Q-(y\sigma \overline{Q_L}Q_R+h.c.) \\
&\hphantom{{\cal L} = -\abs{\partial \sigma}^2-\overline{Q}}\text{ where } \sigma=\frac{f}{\sqrt{2}}e^{i \frac{a}{f}}\\ 
&\supset -\frac{(\partial a)^2}{2}-\overline{Q}\gamma^\mu(\partial_\mu+\frac{yf}{\sqrt{2}}-iA_{\mu}^aT^a) Q + i \frac{y}{\sqrt{2}}a \overline{Q}\gamma_5Q\\
&\xrightarrow[]{Q \text{ triangle loop}} -\frac{a}{32\pi^2f}G\tilde{G} = \frac{i}{32\pi^2}\log(\sigma\vert_{\text{$a$ terms}})G\tilde{G} \ ,
\label{quarkInteg}
\end{aligned}
\end{equation}
where $\sigma$ is a scalar field, $Q_{L,R}$ are left and right handed colored fermions, $T^a$ are the generators of $SU(3)_c$, $A$ is the gluon field of field strength $G$ and $y$ is a Yukawa coupling.

We then see how to generate (\ref{gaugeInvAnom}) from fermions loops, starting from the following lagrangian:\footnote{This procedure, as well as (\ref{higherYukawas}), is uniquely determined by the fermionic gauge charges, see Appendix \ref{KSVZ}.}
\begin{equation}
\begin{aligned}
&{\cal L} \supset \ -y_0\phi_0\overline{Q_{L,0}}Q_{R,0} -\phi_1\overline{Q_{L,1}^{i=1...q}}Y_{1,ij}Q_{R,1}^j \\&\hphantom{{\cal L} \supset \ }-\phi_2\overline{Q_{L,2}^{i=1...q^2}}Y_{2,ij}Q_{R,2}^j+...+h.c.\\
&\xrightarrow[]{\text{triangle loops}} \frac{i}{32\pi^2}\big(\log(\phi_0)+...+q^N\log(\phi_N)\big)G\tilde{G} \\
&\hphantom{\xrightarrow[]{\text{triangle loops}},}= \frac{i}{32\pi^2}\log(\phi_0\phi_1^q...\phi_N^{q^N})G\tilde{G} \ . 
\label{integrateParticles}
\end{aligned}
\end{equation}
This procedure is actually the minimal one (with dimension four Yukawa couplings) that generates an $U(1)_a$\\$\times SU(3)_c^2$ anomaly without generating gauge anomalies (or said differently, that generates (\ref{gaugeInvAnom})). It requires adding colored fermions at each site, in accordance with the fact that (\ref{gaugeInvAnom}) involves all quiver links (and in particular, there is no freedom in using the axion profile to modify the effective scale of the axion-gluons coupling). In terms of couplings defined in (\ref{EandC}), it has ${\cal C}=1$ and ${\cal E}=0$. Note that the lagrangian (\ref{integrateParticles}) (and (\ref{higherYukawas}) below) respects the global symmetry $U(1)_a$. Therefore, it cannot generate the scalar potential (\ref{explicitbreaking}) by quantum corrections. However, since we had to nonetheless include (\ref{explicitbreaking}) as a gravity correction, we should also consider all gauge-invariant non-renormalizable fermionic operators in addition to (\ref{integrateParticles}). Those operators could classically break $U(1)_a$ and generate both the mass of the axion and its couplings to the gluons. In appendix \ref{KSVZ}, we present a model with such fermionic operators.

The number of additional fermions grows exponentially with $N$: for instance, in order to use (\ref{fa}) to bring a Planck scale $f$ down to an intermediate scale $f_a=10^{10-11}$ GeV, we need $\sim q^N \gtrsim 10^{7-8}$ additional fermions (which would however be close to the Planck mass and would thus not spoil gauge coupling unification, or perturbativity far below the Planck mass). Alternatively, if we start with $f$ already at intermediate scale, the strong CP-problem is for instance solved when $f \sim 10^{11} \text{ GeV}, q=3$ and $N=2$. This is enough to ensure the gauge protection according to the discussion following (\ref{protectQCD}), with $1+3+3^2=13$ additional Dirac fermions in the $\textbf{3}$ of $SU(3)_c$. The new fermions spoil asymptotic freedom but keep perturbativity of the strong interactions below the Planck mass. In this specific example, we get a (detectable) coupling to photons from (\ref{EandC}):
\begin{equation}
\begin{aligned}
{\cal L} &\supset (1.7 \times 10^{13} \text{ GeV})^{-1}aF^{\mu\nu}\tilde{F}_{\mu\nu} \ .
\end{aligned}
\end{equation}

\begin{figure}[h]
\centering
\includegraphics[scale=0.42]{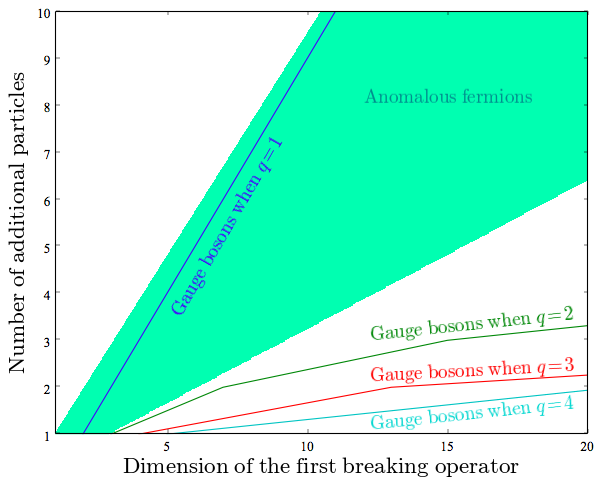}
\caption{Number of additional particles function of \protect\\ the first explicit breaking operator dimension}
\label{protectVSparticles}
\end{figure}

If one wants to circumvent the conclusions of (\ref{integrateParticles}), one can also assign gauge charges to the fermions so that their lowest gauge-invariant mass terms are of higher dimension. One example of this type is 
\begin{equation}
\begin{aligned}
&{\cal L} \supset \ -y_0\phi_0\overline{Q_{L,0}}Q_{R,0} \\&\hphantom{{\cal L} \supset \ }-\frac{1}{M_c^{q+...+q^{N-1}}}\phi_1\phi_2^q...\phi_N^{q^{N-1}}\overline{Q_{L}^{i=1...q}}Y_{ij}Q_{R}^j +h.c.\\
&\xrightarrow[]{\text{triangle loops}} \frac{i}{32\pi^2}\big(\log(\phi_0)+q\log(\phi_1\phi_2^q...\phi_N^{q^{N-1}})\big)G\tilde{G} \\&\hphantom{\xrightarrow[]{\text{triangle loops}},}= \frac{i}{32\pi^2}\log(\phi_0\phi_1^q...\phi_N^{q^N})G\tilde{G} \ ,
\end{aligned}
\label{higherYukawas}
\end{equation}
where $M_c$ is the cutoff of the theory. The action (\ref{higherYukawas}) couples the axion to the gluons via a number of additional fermions independent on $N$, but the high dimension of the second coupling in the first line of (\ref{higherYukawas}) lowers the mass of the $Q_i$ fermions. Since these fermions are colored and unobserved at the LHC, we must impose $m_{Q_i} \gtrsim$ a few TeV, which gives, if one takes as an example $f=\frac{\sqrt{2}}{10}M_c$ and $M_c=M_P$, 
\begin{equation}
\Big(\frac{f}{\sqrt{2}M_c} \Big)^{1+q+...q^{N-1}}M_c \gtrsim \text{TeV} \Rightarrow \frac{q^N-1}{q-1} \lesssim 15 \ .
\label{heavierthanTeV}
\end{equation}
The bound is even more stringent as soon as we decrease $f$ in order to satisfy (\ref{protectQCD}): $f=\frac{\sqrt{2}}{10^6}M_c$ would give $\frac{q^N-1}{q-1} \lesssim 3$. It imposes in particular that we cannot reduce the decay constant of the axion using (\ref{fa}) and (\ref{higherYukawas}) from the Planck scale down to the intermediate scale of invisible axion models. One can interpolate between (\ref{integrateParticles}) and (\ref{higherYukawas}), but then there will either be limitations on $q$ and $N$ due to the high dimension of the mass terms or a number of fermions that grows with $N$ (or both). The situation is illustrated in Figure \ref{protectVSparticles}: even though the use of $q\neq1$ enables the dimension of the first  $U(1)_a$-breaking operator to scale exponentially with respect to the number of quiver groups, the number of fermions necessary to make $U(1)_a$ anomalous scales linearly with respect to this operator dimension (the number of fermions is bounded above by $1+q+...+q^N$, realized in (\ref{integrateParticles}), and we bound it below by $\frac{1+q+...+q^N}{3}$ according to the discussion around (\ref{heavierthanTeV})).

This discussion goes beyond the particular case of the quiver of Figure \ref{quivAxion} and concerns every theory with a protected PQ symmetry: the higher the quality of an accidental PQ symmetry, the higher the number of fermions required to make it anomalous with axionic couplings generated by fermion loops.\footnote{This applies in particular if we enlarge the scalar content of the theory depicted in Figure \ref{quivAxion} to additional scalar fields while keeping the quiver as the main source of protection, as in appendix \ref{DFSZ}.}

Indeed, any axionic coupling term in such a theory free of gauge anomalies must be of the form:
\begin{equation}
i\log({\cal O})G\tilde{G} +h.c. \ ,
\end{equation}
where ${\cal O}$ is by construction gauge-invariant and not invariant under the anomalous global symmetry. If it arises from loops of heavy fermions, it is through the scheme discussed above:
\begin{equation}
{\cal L} \supset -\sum_i({\cal O}_i\overline{\psi_{i,L}}\psi_{i,R})\xrightarrow[]{\text{triangles}}\frac{i}{32\pi^2}\log(\prod_i{\cal O}_i)G\tilde{G}
\end{equation}
(where we assumed that we removed from the sum every pair of vector-like fermions), and ${\cal O}=\prod_i{\cal O}_i$. However, the very notion of accidental axion symmetry means that ${\cal O}$ is an operator of high dimension, so the targeted quality of the axion global symmetry imposes a lower bound on $\text{dim}({\cal O})=\sum_i \text{dim}({\cal O}_i)$, while the definition we adopt for ``heavy" fermions (in our case, unobserved at the LHC) puts an upper bound on $\text{dim}({\cal O}_i)$ for each $i$. The two limits together imply a lower bound on the number of heavy fermions. 

It is useful to analyze how such considerations show up in the other models discussed in section \ref{accidentalQCD}. In the model of \cite{PhysRevD.46.539} one wants to generate $i\log(\phi_0^{q'}\phi_1^{p'})G\tilde{G} +h.c.$ and requires for this $p'+q'$ colored Dirac fermions, whereas in the model of \cite{Hill:2002me,Hill:2002kq} one wants to generate $i\log(\phi_0\phi_1...\phi_N)G\tilde{G} +h.c.$ and requires for this $N+1$ colored Dirac fermions. 

One can now sum up the comparison between those two models and the model under study:
\begin{itemize}
\item in order to protect a QCD axion for a fixed scale of spontaneous breaking $f$, all these models require that the first gauge-invariant $U(1)_{PQ}$-breaking operator ${\cal O}$ be of sufficiently high dimension $d_{\cal O}$ (e.g. $d_{\cal O}=13$ for $f\sim 10^{12}$ GeV). This requires $N\sim d_{\cal O}$ gauge groups for the model of \cite{Hill:2002me,Hill:2002kq}, $p',q'\sim d_{\cal O}$ scalar gauge charges for the one of \cite{PhysRevD.46.539} and only $N\sim\log_{q}(d_{\cal O})$ gauge groups for the model under study
\item the effective decay constant $f_a$ is reduced with respect to the scale $f$ by a factor $\sim d_{\cal O}$ for our model and \cite{PhysRevD.46.539} and by a factor $\sim \sqrt{d_{\cal O}}$ for \cite{Hill:2002me,Hill:2002kq}. This can be understood by studying the $U(1)_a$ charge of ${\cal O}$: if the shift $a\rightarrow a + 2\pi\alpha f_a$ defines the charge normalization, ${\cal O}$ has charge $\sim \sqrt{d_{\cal O}}$ in \cite{Hill:2002me,Hill:2002kq} and $\sim d_{\cal O}$ in \cite{PhysRevD.46.539} and (\ref{gaugeInvAnom}). The higher charge is however due to the high gauge charges in \cite{PhysRevD.46.539} whereas it is due to the clockwork profile of the axion as well as the expression of ${\cal O}$ due to the clockwork gauge charges in our model
\item the number of fermions necessary to generate the axion-gluons coupling is $\sim d_{\cal O}$ in all the models.
\end{itemize}

\section{Axion-like particles}\label{ALPs}

In this section, we study the case of axion-like particles (ALPs), which generically refers to pseudo-Goldstone bosons not designed to solve the strong-CP problem, and whose interactions are consequently less constrained than those of the QCD axion. We will focus on models where the ALP potential is entirely generated by perturbative physics in a UV theory,\footnote{There could also be instantonic contributions to the potential, associated to a confining gauge group with a $U(1)_a$ anomaly. However, since the discussion of section \ref{quarksInt} showed us that making $U(1)_a$ anomalous demands a large number of additional fermions in the theory, especially when $N$ grows, we will for simplicity restrict ourselves to those ALPs which do not have any anomalous couplings.} e.g. gravitational physics, which grants the ALP a small mass even for few quiver sites and which is sufficient to make it a good dark matter candidate. Furthermore, there exist operators which make the dark matter ALP detectable, if for example some standard model particles are charged under the quiver gauge symmetry. This only requires limited additions to the particle content of Figure \ref{quivAxion}, even when the number of quiver sites is large. However, while the $U(1)_a$ protection by the quiver is strong enough to generate quintessence-like mass scales, the need for trans-Planckian field values of usual axion quintessence models is still present in our setup, and is exacerbated by the reduction of the axion effective decay constant.

\subsection{ALPs potentials and dark sector candidates}\label{ALPsDM}

Since most ALPs are used in cosmology, where their treatment can differ significantly from the one of the QCD axion (see \cite{Marsh:2015xka} for a review), let us first discuss the cosmological relevance of our setup. In the non-anomalous setup that we chose to consider in this section, we think of any ALP potential as generated by some classical explicit breaking in a UV theory. The lowest-dimensional gauge-invariant potential of this type for the particle $a$ of (\ref{axProf}) is (\ref{explicitbreaking}), which very weakly breaks $U(1)_a$ and entitle us to call $a$ a pseudo-Goldstone boson, as we discussed in section \ref{GBprotect}. It is a typical periodic potential, consistent with the ALPs origin as a periodic phase degree of freedom, and such potentials are very useful in cosmology: the smallness of the masses and the specific potential they provide make ALPs good dark matter or dynamical dark energy candidates via the misalignment mechanism. The relic density can be calculated once we are given the initial value $a_{\text{init}}$ of the ALP field after inflation and its mass $m_a$ which is taken to be constant in time (and given in (\ref{mass}) for our perturbative setup):
\begin{equation}
\Omega_a \approx \begin{cases}\text{Dark matter: }2\times10^{2}(\frac{m_a}{10^{-22}\text{eV}})^{\frac{1}{2}}(\frac{a_\text{init}}{M_P})^2\\\text{Dark energy: }8\times10^{-2}(\frac{m_a}{10^{-33}\text{eV}})^2(\frac{a_\text{init}}{M_P})^2 \end{cases} .
\label{relic}
\end{equation}
$a_\text{init}$ is given by $a_\text{init}=\epsilon_\text{init} f_a$, where $f_a$ (which defines the periodicity of the ALP potential) is given in (\ref{fa}) and $\epsilon_\text{init}$ depends on one's taste for tuning (we assume in the following that the spontaneous breaking of $U(1)_a$ happens before inflation, see the discussion of appendix \ref{scanAll}). In order for these formulas to be valid, i.e. for the ALP to behave like CDM before radiation-matter equality or like dark energy today, we supplement (\ref{relic}) with:
\begin{equation}
m_a  \begin{cases}\text{DM: }\gtrsim 10^{-28} \text{ eV}\\\text{DE: }\lesssim 10^{-33} \text{ eV} \end{cases} ,
\label{massBounds}
\end{equation}
(where the bound for DM can be pushed up to $m_a\gtrsim 10^{-22} \text{ eV}$ when non-linear cosmological observables are taken into account). 

In our setup, obtaining masses as low as those which appear in (\ref{massBounds}) without tuning is easy (for instance, (\ref{mass}) equals $\sim 10^{-33}$ eV when $f=0.13M_P,q=3,N=4$). However, we can see from the comparison of (\ref{relic}) and (\ref{massBounds}) that axion quintessence demands initial values which are higher than the Planck mass. This can be achieved with some tuning on $\epsilon_\text{init}$ or when the effective decay constant of the axion is increased compared to the mass scales of the model (as in clockwork models which, however, have no mass protection mechanism built in). Since our effective decay constant (\ref{fa}) is reduced, the latter is not an option while the former is not enough to reach the correct energy density (if we insist on keeping $f$ below the Planck mass): indeed if we impose $m_a \lesssim 10^{-33}$ eV, we can only obtain $\Omega_a \ \lesssim 0.05$ and would need at least 13 of such ALPs to reach the observed dark energy density.

In contrast, natural dark matter candidates do arise in our model. In Figure \ref{scanALPs}, we scan the parameters $f$ and $M_c$  for some values of $q$ and $N$ (see appendix \ref{scanAll} for a more complete treatment) which satisfy the condition (\ref{relic}) for $\Omega_{\text{DM}}=0.3$ and (\ref{massBounds}), allowing for $\epsilon_\text{init}$ to range from $0.1$ to $\pi - 0.1$, and allowing a constant multiplying the potential (\ref{explicitbreaking}) ranging from $0.033$ to $30$. In Figure \ref{zoomMP}, we focus on the case where $M_c=M_P$ and on the minimal numbers of quiver gauge groups, in order to highlight the $(m_a,f_a)$ parameter space probed by our model. There, we allow $\epsilon_\text{init}$ to range from $\pi$ to $0.0001$, and we also include the parameter space for the QCD axion (which, due to its temperature-dependent mass, differs for the one of other ALPs).

\begin{figure*} 
\centering
\includegraphics[scale=0.37]{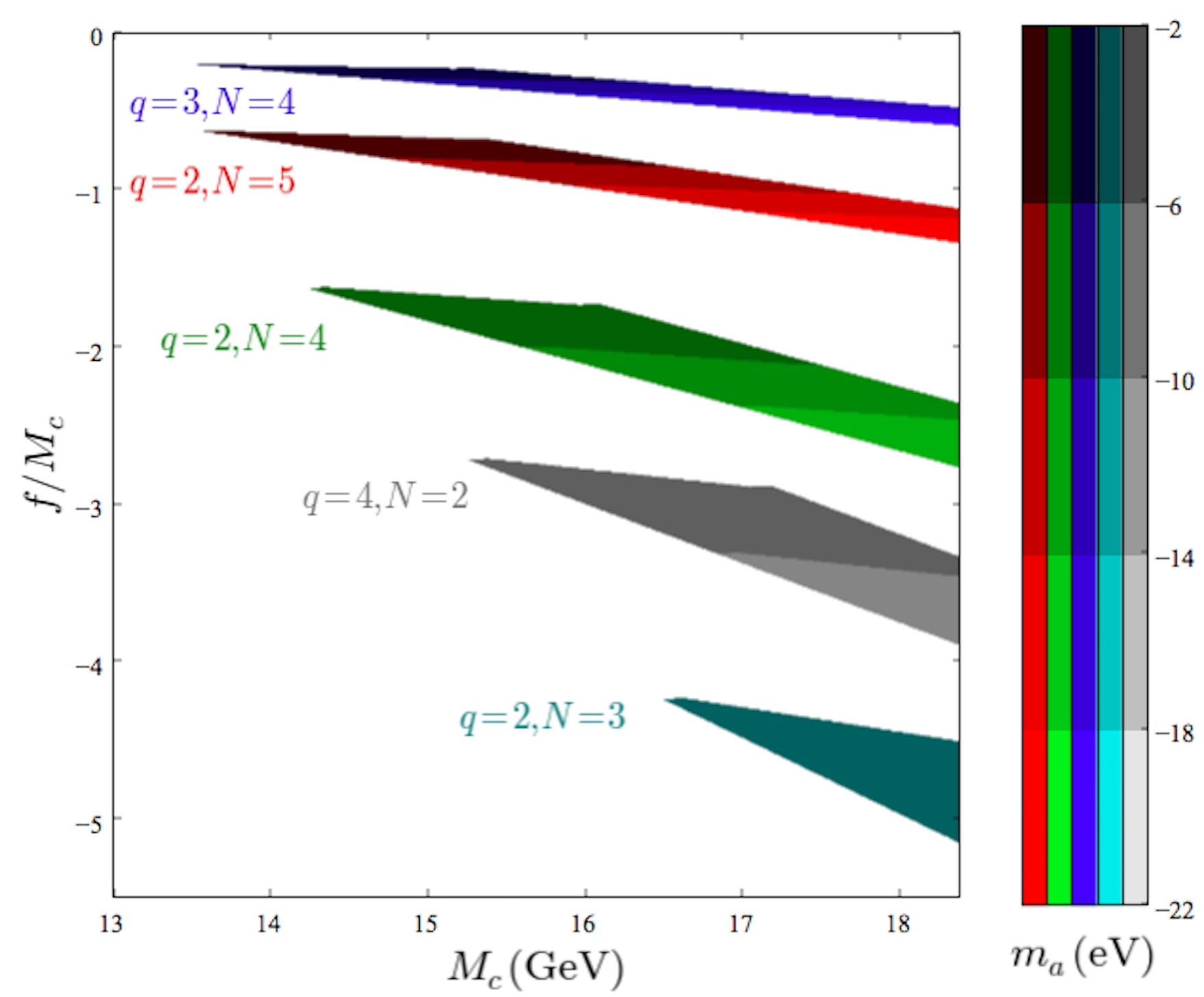}
\caption{Range of parameters for a DM ALP of mass $m_a\leq 10^{-2}$ eV\protect\\(axions saturating the DM relic density are found in colored regions, all axes are log-scale)\protect\\ \vspace{15pt}}
\label{scanALPs}
\includegraphics[scale=0.42]{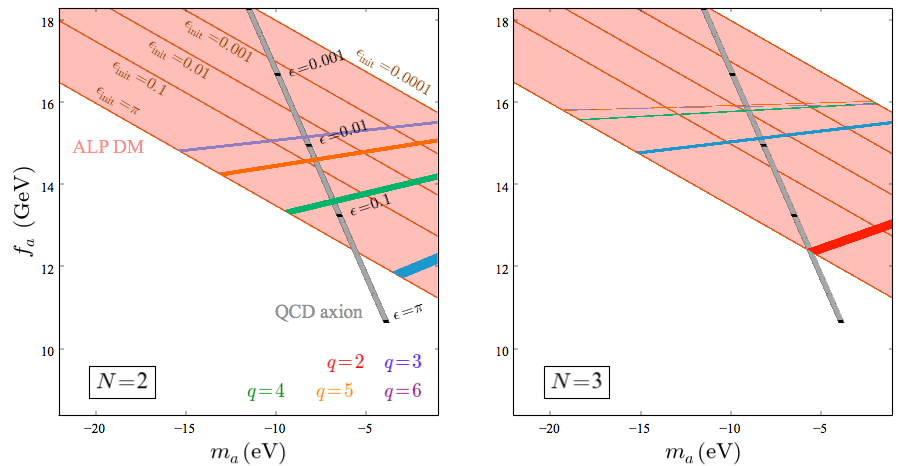}
\caption{Parameter space for a DM ALP of mass $m_a\leq 10^{-2}$ eV when $M_c=M_P$\protect\\(The pink region indicates the parameter space where (\ref{relic}) gives the DM relic density, whereas colored bands show where DM axions are found in our model. The QCD axion parameter space is given by the grey line. Axes are log-scale)}
\label{zoomMP}
\end{figure*}

We see in Figures \ref{scanALPs} and \ref{zoomMP} that we obtain suitable DM candidates, and that the dependence on $q$ and $N$ of the mass (\ref{mass}) allows us to reach very low ALPs masses. These small masses, combined with the high scale $f$ of their associated new physics, are hard to realize in a pure field theoretical framework and are usually thought of as coming from a string axiverse \cite{Svrcek:2006yi,Arvanitaki:2009fg,Cicoli:2012sz}.
Our setup then provides an economical, in the sense of a low number of gauge groups, realization of such values. For instance, the smallest masses discussed in the literature for ultra light dark matter, $m_a \sim 10^{-21}-10^{-22}$ eV, require $M_c \approx M_P$ and are obtained for $f \approx 0.2M_P$, $q=3$ and $N=4$ (for the choices of $q$ and $N$ displayed, to be compared with $p',q'$ or $N \sim 120-130$ respectively for \cite{PhysRevD.46.539} and \cite{Hill:2002me,Hill:2002kq}, discussed in sections \ref{accidentalQCD} and \ref{quarksInt}). This example, as well as Figure \ref{zoomMP}, shows that, even though we scan different values of $M_c$ in Figure \ref{scanALPs}, a gravitational origin ($M_c=M_P$) for (\ref{explicitbreaking}) is sufficient to reproduce the cosmological relic density of dark matter.

In order to conclude that such ALPs are to play a role in the cosmic evolution, we must check that their lifetime can be comparable to or bigger than the age of the universe. In generic models, there is a decay channel of an ALP into two photons, usually coming from a $U(1)_{PQ} \times U(1)_{em}^2$ anomaly. Even though there is no anomaly in the models of this section, non-anomalous, CP-even and gauge invariant operators that enable this decay exist. For instance, they can arise if we couple one of the quiver sites of Figure \ref{quivAxion} to an anomaly-free set of electrically charged fermions displayed in Table \ref{ALPtogamma}, while we keep the standard model particles uncharged under the quiver gauge group.
\begin{table}[h]
\centering
\begin{tabular}{ccccc} 
  \hline
&$U(1)_i$&$U(1)_{i+1}$&$U(1)_{em}$&$U(1)_a$\\
\hline
$\psi_{L,1}$&$-1$&$0$&$ne$&$q_1$\\
\hline
$\psi_{R,1}$&$0$&$-q$&$ne$&$q_1+q^i$\\
\hline
$\psi_{L,2}$&$1$&$0$&$-ne$&$q_2$\\
\hline
$\psi_{R,2}$&$0$&$q$&$-ne$&$q_2-q^i$\\
\hline
\end{tabular}
\caption{Anomaly-free set of fermions coupling the ALP to the photon field\protect\\
(the three first columns indicate the gauge charges of the fields whereas the last one gives the PQ charges induced by (\ref{exampleAGG}), as functions of $q_1$ and $q_2$ which are arbitrary)}
\label{ALPtogamma}
\end{table}

With such charges, one can write Yukawa couplings $y_{1,2}$ to $\phi_i$:
\begin{equation}
{\cal L} \supset -y_1\phi_i\overline{\psi_{R,1}}\psi_{L,1} - y_2\phi_i\overline{\psi_{L,2}}\psi_{R,2}+h.c. \ .
\label{exampleAGG}
\end{equation}
The effective operators describing the decay $a \rightarrow \gamma \gamma$ then are (see Appendix \ref{calculation} for the computation):
\begin{equation}
\ba
&{\cal L} \supset \frac{n^2e^2q^i}{192\pi^2\sqrt{1+q^2+...q^{2N}}f}\Big(\frac{1}{m_1^2}-\frac{1}{m_2^2}\Big)\\&\hphantom{{\cal L} \supset \frac{n^2e^2q^in^2e^2}{a}}
\times(-\Box a F\tilde{F}+2\partial_\mu a F_{\nu\eta}\partial^\eta\tilde{F}^{\mu\nu})  \ ,
\ea
\label{atogamma}
\end{equation}
where $m_1=\frac{y_1f}{\sqrt{2}},m_2=\frac{y_2f}{\sqrt{2}}$ and $F$ is the photon field strength. Notice that, contrary to anomalous couplings such as (\ref{gaugeInvAnom}), non-anomalous interactions are {\it site} {\it localized}, and exhibit clockwork-like effects due to the profile (\ref{axProf}). This feature will be present in all the operators discussed in this section. We can also see that the non-anomalous nature of the ALP-photons coupling makes this interaction of derivative type and of higher dimension than usual anomalous $aF\tilde{F}$ terms, so this decay does not make our ALPs unstable over the cosmic history. Indeed these couplings give a decay rate:
\begin{equation}
\Gamma_{a\rightarrow \gamma \gamma}=\frac{q^{2i}n^4\alpha^2m_a^7}{1024\pi^3(1+q^{2}+...q^{2N})f^2}\Big(\frac{1}{m_1^2}-\frac{1}{m_2^2}\Big)^2 \ , 
\label{decay2}
\end{equation}
where $\alpha$ is the fine structure constant. Hence, we conclude that this decay channel is harmless with respect to the cosmic evolution of our ALPs.\footnote{For ultra-light dark matter with $m\sim10^{-21}$ eV, if we choose $f=m_1=2m_2=0.3M_P$ and $n=1$, the decay rate is $\sim \frac{q^{2i}}{1+q^{2}+...q^{2N}}(10^{-300}s^{-1})$.} Indeed, as guessed above, the non-anomalous nature of the ALP-photons coupling forces the $m_a$ factor to appear in the decay rate (\ref{decay2}) at a higher power than in the case of usual $aF\tilde{F}$-induced decays and ensures a long ALP lifetime. The clockwork-like dependence of (\ref{decay2}) only tends to weaken the ALP couplings to photons when matter is coupled to the first quiver sites. 

The example of Table \ref{ALPtogamma} is a realization of the more general gauge-invariant non-anomalous operators, coupling the axion to the photon field, which we can write within the effective field theory:
\begin{equation}
\begin{aligned}
&\frac{1}{\Lambda^4}D^\mu D_\mu \phi_i\phi_i^* F^{\mu\nu} \tilde{F}_{\mu\nu}, \frac{1}{\Lambda^4}D_\mu \phi_i\phi_i^* \partial^\eta F^{\mu\nu}  \tilde{F}_{\eta\nu} \\&\hphantom{\frac{1}{\Lambda^4}D^\mu D_\mu \phi_i\phi_i^* F^{\mu\nu} \tilde{F}_{\mu\nu}andan}\text{ and } \frac{1}{\Lambda^4}D_\mu \phi_i\phi_i^* \partial^\eta \tilde{F}^{\mu\nu}  F_{\eta\nu}\\
&\xrightarrow[]{\text{terms linear in $a$}}  \frac{iq^if}{2\sqrt{1+...q^{2N}}\Lambda^4}\big(\Box a F\tilde{F},\partial_\mu a \partial^\eta F^{\mu\nu}  \tilde{F}_{\eta\nu} \\&
\hphantom{\xrightarrow[]{\text{terms linear in $a$}} \frac{iq^if}{2\sqrt{1+...q^{2N}}\Lambda^4}(}\text{ and } \partial_\mu a \partial^\eta \tilde{F}^{\mu\nu}  F_{\eta\nu}\big) \ ,
\label{agammagamma}
\end{aligned}
\end{equation}
where $D$ is the covariant derivative, there is no summation over the index $i$ and $\Lambda$ the scale at which this operator is generated. For instance, in the example of Table \ref{ALPtogamma} $\Lambda$ is equal to the mass of the $\psi$ fermions. Since (\ref{agammagamma}) preserves $U(1)_a$, $\Lambda$ does not have to be equal to $M_c$ which was the scale of classical explicit breaking, even though there could also be $U(1)_a$ preserving interactions at scale $M_c$ (for instance there could be gravitational contributions of the form (\ref{agammagamma}) where $\Lambda=M_P$). Thus, in a minimal, agnostic approach, we should consider effective theory operators such as (\ref{agammagamma}), supplemented by the potential (\ref{explicitbreaking}) where three independent scales are used: the scale $f$, and $f_a$ which follows, which are the scales of spontaneous breaking of $U(1)_a$, are given by the quiver and the renormalizable scalar potential in (\ref{model}). The scale $M_c$, at which $U(1)_a$ is explicitly broken, must verify $M_c > f$ for the effective lagrangian to be valid and $M_c \lesssim M_P$ since gravity anyway breaks $U(1)_a$. Finally, $\Lambda$ is a scale of additional physics which generates couplings of the quiver fields to other sectors of the theory, like the SM. It must respect $\Lambda \gtrsim f$, since the new physics can lie at (almost) scale $f$, like in the example of Table \ref{ALPtogamma}, but should not be at a lower scale than the effective theory one.

From (\ref{agammagamma}) we can calculate the decay rate of an axion into two photons:
\begin{equation}
\Gamma_{a\rightarrow \gamma \gamma} \sim \frac{q^{2i}m_a^7f^2}{16\pi(1+q^{2}+...q^{2N})\Lambda^8} \ .
\label{decay}
\end{equation}
This result generalizes (\ref{decay2}) and of course does not spoil the conclusions made with $\Lambda \sim f$ since the dependence on $m_a$, responsible of the low value of $\Gamma_{a\rightarrow \gamma \gamma}$, has not changed and $\Lambda \gtrsim f$ can only weaken the decay rate.

\subsection{Detection via NMR}\label{NMR}

The non-anomalous couplings to photons of (\ref{agammagamma}), too weak to destabilize the cosmic history of our ALPs, are also too weak to be probed by current ALPs searches, which rely on a dimension 5 anomalous $aF\tilde{F}$ coupling. Non-anomalous dimension 5 generic couplings of a Goldstone pseudoscalar $a$ to a detector's matter can be written \cite{Srednicki:1985xd} as follows:
\begin{equation}
\frac{g_{aee}}{f_a} \partial_\mu a \ \overline{e}\gamma^\mu\gamma_5e \text{ and } \frac{g_{aNN}}{f_a} \partial_\mu a \ \overline{N}\gamma^\mu\gamma_5N \ ,
\end{equation}
where $g$'s are dimensionless coupling constants, $f_a$ is again the axion decay constant, and $N$ and $e$ are respectively the nucleon and electron fields. In our setup, they can be generated in field theory if we charge the first family of the standard model under $U(1)_{i,i+1}$, according to Table \ref{Spinmodel},\footnote{All anomalies involving at least one standard model factor are canceled with these charges. One must however add additional fermions only charged under $U(1)_{i,i+1}$ to cancel the $U(1)_{i,i+1} \times U(1)_{i,i+1} \times U(1)_{i,i+1}$ and $U(1)_{i,i+1}$-gravity anomalies. See appendix \ref{KSVZ} for explicit examples.} in a way which gives them $U(1)_a$ charges.
\begin{table}[h]
\begin{equation*}
\begin{array}{ccccccc}
\hline
\text{Fields}&SU(3)&SU(2)&U(1)_Y&U(1)_i&U(1)_{i+1}&U(1)_a\\
\hline
Q_L&\textbf{3}&\textbf{2}&\dfrac{\strut 1}{\strut 6}&0&0&q_Q\\
\hline
u_R&\textbf{3}&\textbf{1}&\dfrac{\strut2}{\strut3}&-q&1&q_Q+q_H+q^i\\
\hline
d_R&\textbf{3}&\textbf{1}&-\dfrac{\strut1}{\strut3}&q&-1&q_Q-q_H-q^i\\
\hline
L_L&\textbf{1}&\textbf{2}&-\dfrac{\strut1}{\strut2}&0&0&q_L\\
\hline
e_R&\textbf{1}&\textbf{1}&-1&q&-1&q_L-q_H-q^i\\
\hline
H&\textbf{1}&\textbf{2}&-\dfrac{\strut1}{\strut2}&0&0&q_H\\
\hline
\end{array}
\end{equation*}
\caption{SM charges that produce an ALP-spin coupling\protect\\
(the first columns indicate the gauge charges of the fields whereas the last one gives the PQ charges induced by (\ref{lagSpin}), as functions of $q_Q,q_H$ and $q_L$ which are arbitrary)}
\label{Spinmodel}
\end{table}
At lowest order, the most general lagrangian is the SM lagrangian where only the first family Yukawa terms have been modified:
\begin{equation}
\begin{aligned}
{\cal{L}} \supset& -\frac{1}{M_c}\Big(\overline{u_R}H\phi_iY_uQ_L + \overline{d_R}(H\phi_i)^*Y_dQ_L \\
&\hphantom{-\frac{1}{M_c}\Big(\overline{u_R}H\phi_iY_uQ_L++} +\overline{e_R}(H\phi_i)^*Y_eL_L\Big) +h.c.\\
\supset& -\frac{vf}{2M_c}\Big(\overline{u}[e^{i\frac{q^ia}{\sqrt{1+...q^{2N}}f}\gamma_5}Y_u]u \\&
+\overline{d}[e^{-i\frac{q^ia}{\sqrt{1+...q^{2N}}f}\gamma_5}Y_d]d+\overline{e}[e^{-i\frac{q^ia}{\sqrt{1+...q^{2N}}f}\gamma_5}Y_e]e\Big) \ ,
\end{aligned}
\label{lagSpin}
\end{equation}
where $v$ is the Higgs vev and where we assumed that these higher order Yukawa couplings come from the same physics which generated (\ref{explicitbreaking}), even though the fact that the precise scale $M_c$ divides these operators is of no importance for what follows. One can make the fermion masses in (\ref{lagSpin}) real with an appropriate chiral redefinition of the fermions, and obtain from their kinetic terms the expected couplings:
\begin{equation} 
{\cal{L}} \supset \ \frac{-iq^i \partial_\mu a}{2\sqrt{1+...+q^{2N}}f}(\overline{u}\gamma_5\gamma^\mu u +\overline{d}\gamma_5 \gamma^\mu d+\overline{e}\gamma_5 \gamma^\mu e) \ , 
\label{Spincouplings}
\end{equation}
where no anomalous term appeared since the $U(1)_a$ symmetry is anomaly-free and where we note that, similar to what was observed previously for axion-photons couplings, the ALP-spin coupling of (\ref{Spincouplings}) is site dependent due to the clockwork profile (\ref{axProf}). 

\begin{figure}[h]
\includegraphics[scale=0.4]{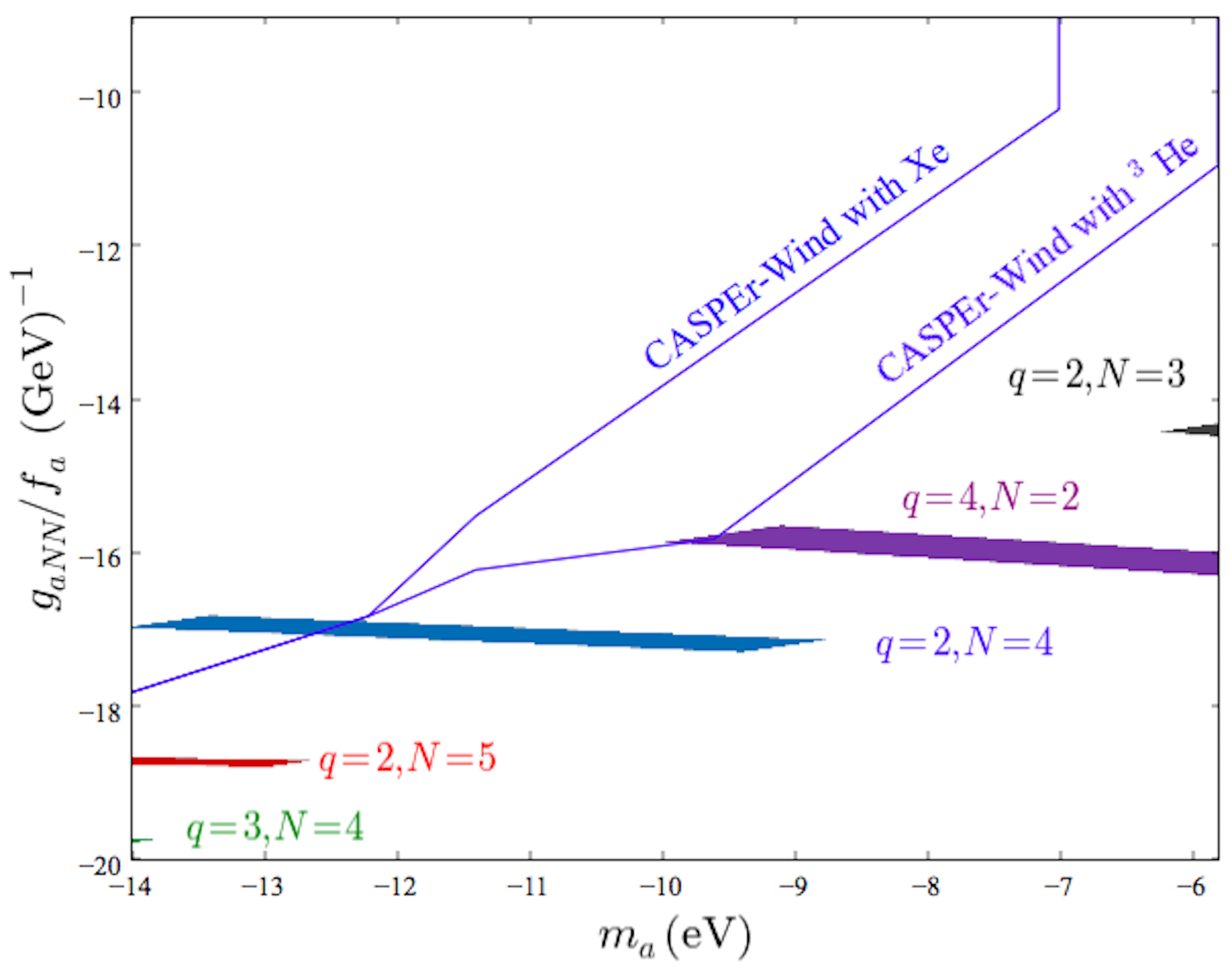}
\caption{Sensitivity of CASPEr-Wind to the ALPs\protect\\(colored regions indicate axions suitable to saturate the DM relic density, blue curves set the limit of the upper left part of the plot where the sensitivity of CASPEr-Wind allows for a DM detection. Both axes are log-scale)}
\label{CasperSensitivity}
\end{figure}

If the mass (\ref{mass}) of the ALP is such that it constitutes part of the dark matter, these couplings may soon be tested via Nuclear Magnetic Resonance\footnote{Bounds already exist on axion-mediated spin-dependent forces between particles, but they do not constrain models with high or intermediate scale axion decay constants.} (NMR) by the CASPEr-Wind experiment \cite{Graham:2013gfa}. As an illustration, in Figure \ref{CasperSensitivity} we assume that the coupling (\ref{Spincouplings}) is located at site $i=0$ of the quiver and restrict ourselves to the $(q,N)$ values displayed in Figure \ref{scanALPs} and to the gravitational breaking of the axionic symmetry (i.e. to the case where $M_c=M_P$, see appendix \ref{scanAll} for a more general study). We then see that CASPEr-Wind can detect some of the ALPs discussed in this paper (one example is for $f \lesssim 5\times10^{15} \text{ GeV}$, $q=2$ and $N=4$). Thus the present model, while invisible to experiments based on axion-photons couplings, can be probed and constrained by NMR-based ALPs searches. Note however that, in order for (\ref{lagSpin}) to be consistent with the observed values of the fermion masses, there should not be a too strong hierarchy between $f$ and $M_c$. Notice also that possible FCNC effects induced by such Yukawa couplings (see for ex. \cite{Marques-Tavares:2018cwm}) are completely unobservable due to the high values of $f$ and $M_c$. 

Like we did in (\ref{agammagamma}), we can generalize such couplings in the gauge-invariant effective theory:
\begin{equation}
\begin{aligned}
&\frac{1}{\Lambda^2}D_\mu \phi_i\phi_i^* \ \overline{e}\gamma^\mu\gamma_5e \text{ and } \frac{1}{\Lambda^2}D_\mu \phi_i\phi_i^*\ \overline{N}\gamma^\mu\gamma_5N \\
&\xrightarrow[]{\text{terms linear in $a$}} \ \frac{iq^if}{2\sqrt{1+...q^{2N}}\Lambda^2}(\partial_\mu a \ \overline{e}\gamma^\mu\gamma_5e \text{ and }\\
&\hphantom{\xrightarrow[]{\text{terms linear in $a$}} \ \frac{iq^if}{2\sqrt{1+...q^{2N}}\Lambda^2}(\partial_\mu+} \partial_\mu a \ \overline{N}\gamma^\mu\gamma_5N) \ ,
\end{aligned}
\label{axionSpin}
\end{equation}
Once again, the scale $\Lambda$ is a priori undetermined since (\ref{axionSpin}) does not break $U(1)_a$, for instance $\Lambda=M_P$ if (\ref{axionSpin}) is of gravitational origin. However, taking $\Lambda=M_P$ does not allow to detect our DM candidates, contrary to the case of (\ref{Spincouplings}) where it is equal to $f$. 

\section{Conclusions}
  
A generic problem for QCD axion models or models for  ultra-light PGB's as candidates for DM or quintessence is to control non-anomalous contributions to  the PGB potential coming from classical explicit breaking of the global symmetry, e.g by gravitational interactions.  Such contributions  would shift the axion field in the minimum to unacceptably large for solving the strong CP problem values  and could jeopardize the possibility of having ultralight axion-like particles  as DM candidates. This  problem has been often addressed in the literature in models with the global symmetry being an accidental remnant of gauge  symmetries. 
Another issue  is the origin  of the scale $f_a$ (the axion decay constant) and of the potential hierarchy $f_a\ll M_P$. Such a hierarchy is required for the generic  QCD axion window  but not needed for the ALPs as the dark matter candidates only, or even not acceptable for $m_\text{ALPs}\sim O(10^{-15}-10^{-20})$ eV. One more difference  between  the QCD axion and the ALPs models is that the former requires a set of colored fermions to generate the anomalous coupling to gluons. 

In this paper we have studied the  QCD axion or cosmological ALPs in a model
inspired by the recent interest in 4-dimensional clockwork models, with the global symmetry accidentally arising due to gauge symmetries. For the QCD axion we have analyzed the connection between the degree of protection of the axion mass against gravitational corrections, the explanation of the hierarchy $f_a\ll M_P$ and the number of colored fermions needed to generate anomalous couplings to gluons, all linked together by the underlying gauge symmetries. In the DM ALPs models, assuming that their mass is solely given by gravitational corrections, we have identified the parameter space such that the scale $f_a$ and the mass $m_a$ combine to give the observed relic abundance. In the latter case, we have used gravitational corrections in a constructive way, instead of invoking new anomalous gauge interactions as a source of properly adjusted explicit breaking and ignoring non-anomalous gravitational contributions.

Based on that model and on the comparison  with earlier models in the literature, we have derived certain general conclusions on the QCD axion models that use global symmetries as consequences of gauge symmetries, to protect the PGB potential against large non-anomalous explicit breaking. In such models the scale $f$ of spontaneous global symmetry breaking is not identical to the axion decay constant $f_a$, with $f/f_a >1$. The larger the ratio $f/f_a$ the better the protection but sufficient protection is obtained already for $f/f_a\sim {\cal O}(10)$.  Furthermore, the number of colored fermions needed to generate axion anomalous couplings is approximately equal  to the ratio $f/f_a$. Thus, the minimal sufficient protection puts the lower bound $\geq {\cal O}(10)$ on the number of new colored  fermions.

Several results for the QCD axion are specific for our scenario based on an abelian gauge theory quiver with scalar bifundamental fields. The contributions from non-anomalous explicit breaking
effects to the axion  potential, and in particular to its mass are a function of the gauge charge assignment $(1,-q)$ for the scalars and  the number of quiver sites. Already with $q>1$ but $q={\cal O}(1)$ and  a few quiver sites ($N={\cal O}(1)$, for instance $q=3$ and $N=2$), the mass protection against gravitational effects  is sufficiently strong, with $f/f_a\sim {\cal O}(10)$.
The number of heavy colored fermions is growing approximately exponentially with the  number $N$ of gauge groups as $q^N$. Whereas a large number of sites $N$ is not needed for the mass protection of the PGB, it could be an useful option in order to decrease the axion decay constant from a large (Planck or string) value to an intermediate scale, since qualitatively $f_a \sim f/q^N$. Notice that the heavy fermions masses necessary to generate the axion-gluons coupling can be close to the Planck scale, without creating problems with the perturbativity of the low-energy theory even for large $q^N$. The minimal number of sites $N$ in order to realize realistic models of this type is then way lower than for $q=1$. Such a high N also connects the 4d model to the deconstruction of a five-dimensional abelian vector model on a linear dilaton background, with Dirichlet boundary conditions for the 4d components of the gauge field, which shares the same low-energy limit as the 4d theory. This gives some intuition to understand some features of the 4d model.

We have shown that the clockwork inspired model is a particularly economical model for a very light ALP as a DM candidate, with the observed relic abundance. Interestingly, 
a small number $N$ of gauge groups is  required for gravitational corrections to induce a just right ALP potential, without referring to any additional strongly interacting sector and its chiral anomalies. 
Such a dark matter axion-like particle can be coupled to the standard model with a small number of extra particles, if any, that does not depend on  $N$. In particular, those couplings would be generated at a given site of the quiver and be sensitive to the clockwork profile of the axion. Such models can be tested via Nuclear Magnetic Resonance experiments, which record the matter spin precession due to the oscillation of the dark matter field. Pseudo-Goldstone quintessence models of dynamical dark energy can also be obtained in such a setup, but their construction faces usual challenges, such as a trans-Planckian axion decay constant, in order to recover the observed energy density.

\section*{Acknowledgements}
Q.B. and E.D. would like to thank Debtosh Chowdhury for helpful discussions. Q.B. and E.D. acknowledges partial support from the ANR Black-dS-String. The work of SP is partially supported by the National Science Centre, Poland, under research grants DEC-2014/15/B/ST2/02157,  DEC-2015/18/M/ST2/00054 \\and  DEC-2016/23/G/ST2/04301.

\begin{appendices}

\section{5d deconstruction on a linear dilaton background}\label{5d}

\subsection{Abelian gauge field}

We recall in this appendix the link between the 5 dimensional deconstructed theory of an abelian gauge field on a linear dilaton background and the low-energy modes of the 4d model defined in eq. (\ref{model}). 

We start by considering a 5d manifold which is the product of the 4d Minkowski space with an interval of length $L$, with a 5d theory living on it:
\begin{equation}
{\cal S} = \int d^5z \sqrt{-g} \ {\cal L}(z^M) = \int d^4x \int_0^L dy \sqrt{-g} \ {\cal L}(x^\mu,y) \ ,
\end{equation}
where we split the 5d coordinates $z^M$ into 4d Minkowski coordinates $x^\mu$, and the position along the interval $y$.

We discretize the fifth dimension interval down to a regularly-spaced lattice of $N+2$ sites. Defining $\Delta_4=\frac{L}{N+1},y_i=i\Delta_4$ (where $i$ runs from $0$ to $N+1$), this amounts to replacing:
\begin{equation}
\begin{aligned}
\int_0^L dy\ f(x,y) &\rightarrow \sum_{i=0}^{N}\Delta_4 f(x,y_i) \ ,\\
\partial_y f(x,y_i) &\rightarrow \frac{f(x^\mu,y_{i+1})-f(x,y_i)}{\Delta_4} \ .
\end{aligned}
\label{lattice}
\end{equation}
We choose to denote $f(x,y_i)=f_i(x)$ in what follows. We do not wish to study the dynamics of the background and restrict to the following static metrics:
\begin{equation}
ds^2=g_{MN}dz^Mdz^N=e^{-2a(y)}(\eta_{\mu\nu}dx^\mu dx^\nu+e^{2b(y)}dy^2) \ .
\label{metric}
\end{equation}
The case $a=ky,b=0$ describes the so-called linear dilaton background in the conformally flat frame, whereas $a=b=ky$ is the Randall-Sundrum metric.

We will study a five-dimensional abelian theory of lagrangian
\begin{equation}
\begin{aligned}
\sqrt{-g}{\cal L}&=\sqrt{-g}\Big(-\frac{1}{4}g^{MP}g^{NQ}F_{MN}F_{PQ}\Big)\\&= -\frac{e^{-5a+b}}{4}\Big(e^{4a}F^{\mu\nu}F_{\mu\nu}+2e^{4a-2b}F^{\mu4}F_{\mu4}\Big) \\
&= -\frac{e^{-a+b}}{4}F_{\mu\nu}^2-\frac{e^{-a-b}}{2}(\partial_\mu A_4 - \partial_4 A_\mu)^2 \ ,
\end{aligned}
\end{equation}
where the 4d indices are contracted using the Minkowski metric. We impose 5d Dirichlet boundary conditions for $A_\mu$ and Neumann conditions for $A_4$ :
\begin{equation}
\begin{aligned}
&A_\mu(x,y=0,L)=0 \ , \\
&\partial_4 (e^{-a-b}A_4)(x,y=0,L)=0 \ .
\end{aligned}
\label{boundary}
\end{equation}
Deconstruction now yields:
\begin{equation}
\begin{aligned}
&\int_0^L dy \ \sqrt{-g}{\cal L}\\&=\sum_{i=0}^{N}\Delta_4 \Big( -\frac{e^{-a_i+b_i}}{4}F_{i,\mu\nu}^2 \\&\hphantom{\sum_{i=0}^{N}\Delta_4 \Big( -++}- \frac{e^{-a_i-b_i}}{2}\Big[\partial_\mu A_{i,4} - \frac{A_{i+1,\mu}-A_{i,\mu}}{\Delta_4}\Big]^2 \Big)\\
&= \sum_{i=0}^{N}\Big( -\frac{1}{4}F_{i,\mu\nu}^{'2}\\& \ - \frac{1}{2}\Big[\partial_\mu A'_{i,4} - \frac{e^{\frac{a_{i+1}-a_{i}-b_{i+1}-b_{i}}{2}}A'_{i+1,\mu}-e^{-b_i}A'_{i,\mu}}{\Delta_4}\Big]^2 \Big)  \ ,
\end{aligned}
\label{4dlag}
\end{equation}
where we defined $A'_{i,\mu}=e^{\frac{-a_i+b_i}{2}}\sqrt{\Delta_4}A_{i,\mu}$ (with $F'$ its associated field strength) and $A'_{i,4}=e^{\frac{-a_i-b_i}{2}}\sqrt{\Delta_4}A_{i,4}$. Dropping the primes and using the boundary conditions, we finally obtain
\begin{equation}
\begin{aligned}
&{\cal L}= -\frac{1}{4}\sum_{i=1}^{N}F_{i,\mu\nu}^2 \\&- \frac{1}{2}\sum_{i=1}^{N-1}\Big(\partial_\mu A_{i,4} - \frac{e^{\frac{a_{i+1}-a_{i}-b_{i+1}-b_{i}}{2}}A_{i+1,\mu}-e^{-b_i}A_{i,\mu}}{\Delta_4}\Big)^2\\
&- \frac{1}{2}\Big(\partial_\mu A_{0,4} - \frac{e^{\frac{a_{1}-a_{0}-b_{1}-b_{0}}{2}}A_{1,\mu}}{\Delta_4}\Big)^2 \\&- \frac{1}{2}\Big(\partial_\mu A_{N,4} + \frac{e^{-b_{N}}A_{N,\mu}}{\Delta_4}\Big)^2 \ .
\end{aligned}
\end{equation}
Specializing to the linear dilaton background for which $a=ky,b=0$, the lattice action now becomes
\begin{equation}
\begin{aligned}
{\cal L}=& -\frac{1}{4}\sum_{i=1}^{N}F_{i,\mu\nu}^2 - \frac{1}{2}\sum_{i=1}^{N-1}\Big(\partial_\mu A_{i,4} - \frac{e^{\frac{k}{2}}A_{i+1,\mu}-A_{i,\mu}}{\Delta_4}\Big)^2\\
&- \frac{1}{2}\Big(\partial_\mu A_{0,4} - \frac{e^{\frac{k}{2}}A_{1,\mu}}{\Delta_4}\Big)^2 - \frac{1}{2}\Big(\partial_\mu A_{N,4} + \frac{A_{N,\mu}}{\Delta_4}\Big)^2 \ ,
\end{aligned}
\label{deconstructed}
\end{equation}
where we made the replacement $k \rightarrow \frac{k}{\Delta_4}$. 
Defining $q=e^{\frac{k}{2}},f=\frac{1}{\Delta_4},\phi_i=\frac{f}{\sqrt{2}}e^{-i\frac{A_{i,4}}{f}},D_{\mu}\phi_i=\big(\partial_{\mu}-i(1-\delta_{i,0})A_{\mu,i}+iq(1-\delta_{i,N})A_{\mu,i+1}\big)\phi_i$, we can rewrite (\ref{deconstructed}) as
\begin{equation}
{\cal L}=-\frac{1}{4}\sum_{i=1}^{N}F_{\mu\nu,i}F^{\mu\nu}_i-\sum_{k=0}^{N}\abs{D_{\mu}\phi_k}^2 \ ,
\label{UVcompleted}
\end{equation}
thus establishing the link between the low-energy limit of the 4d model of section \ref{4d} and the deconstruction on a linear dilaton background of an 5d abelian vector mode with boundary conditions (\ref{boundary}).

Finally, the Wilson line $e^{i\int dy \ A_4(x,y)}$ gets mapped to the $U(1)_a$-violating potential of (\ref{gaugeInvOp}):
\begin{equation}
e^{i\int dy \ \sqrt{\Delta_4}A_4(x,y)} = e^{i \sum_{i=0}^{N} q^i\Delta_4 A'_{i,4}(x)}= \frac{\phi_0\phi_1^q...\phi_N^{q^N}}{(\frac{f}{\sqrt{2}})^{1+q+...q^N}} \ . 
\end{equation}

\subsection{Charged bulk fermion}

Deconstructed fermions might be useful in order to get insights on how $U(1)_a$ can be made anomalous or classically broken \cite{Hill:2002me,Hill:2002kq}. However, as we will see below, this procedure is not applicable in our setup. Indeed, let us consider the action of a bulk fermion charged under the abelian symmetry of the previous section:
\begin{equation}
\begin{aligned}
\sqrt{-g}{\cal L}=& \sqrt{-g} \Big(-\frac{1}{2}\overline{\Psi}[\gamma^M(\partial_M-ie_4A_M)+m]\Psi+h.c.\Big)\\
=& e^{-5a+b} \Big(-\frac{1}{2}\overline{\Psi}[e^{a}\gamma^\mu(\partial_\mu-ie_4A_\mu)+m]\Psi\\
&\hphantom{e^{-5a} -\frac{1}{2}\overline{\Psi}}-\frac{e^{a-b}}{2}\overline{\Psi}\gamma^4[\partial_4-ie_4A_4]\Psi+h.c.\Big)\\
=& e^{-4a+b} \Big(-\frac{1}{2}\overline{\Psi}[\gamma^\mu(\partial_\mu-ie_4A_\mu)+e^{-a}m]\Psi\\
&\hphantom{e^{-5a+b} -\overline{\Psi}}-\frac{e^{-b}}{2}\overline{\Psi}\gamma^4[\partial_4-ie_4A_4]\Psi+h.c.\Big)\ ,
\end{aligned}
\end{equation}
where we did not include the spin connection of the metric (\ref{metric}), calculable from the vielbein $e_A^M=\delta_A^M\times$ $(e^{a-b\delta_5^M})$, since it cancels out in the action, and $\gamma^4$ can be taken equal to the 4d $\gamma_5$. Deconstructing, using the normalized bosonic fields and defining $\Psi_i'=\sqrt{\Delta_4}e^{-2a_i+\frac{b_i}{2}}\Psi_i$ we get:
\begin{equation}
\begin{aligned}
\int dy& \ \sqrt{-g}{\cal L} \rightarrow\\ \sum_{i=0}^N& \Delta_4 e^{-4a_i+b_i} \Big(-\frac{1}{2}\overline{\Psi_i}[\gamma^\mu(\partial_\mu-ie_4A_{i,\mu})+e^{-a_i}m]\Psi_i\\
&-\frac{e^{-b_i}}{2}\overline{\Psi_i}\gamma_5[\frac{\Psi_{i+1}-\Psi_i}{\Delta_4}-ie_4A_{i,4}\Psi_{i+1}]+h.c.\Big)\\
= \sum_{i=0}^N& \Big(-\frac{1}{2}\overline{\Psi_i'}[\gamma^\mu(\partial_\mu-i\frac{e^{\frac{a_i-b_i}{2}}e_4}{\sqrt{\Delta_4}}A_{i,\mu}')+e^{-a_i}m]\Psi_i'\\
&-\frac{1}{2}\overline{\Psi_i'}\gamma_5[\frac{e^{2a_{i+1}-2a_i-\frac{b_{i+1}}{2}-\frac{b_{i}}{2}}\Psi_{i+1}'-e^{-b_i}\Psi_i'}{\Delta_4}\\&\hphantom{+++}-i\frac{e_4}{\sqrt{\Delta_4}}e^{2a_{i+1}-\frac{3a_i}{2}-\frac{b_{i+1}}{2}}A_{i,4}'\Psi_{i+1}']+h.c. \Big)\ .
\end{aligned}
\end{equation}
We now restrict the discussion to the linear dilaton background, with the vector boundary conditions of the previous section,  to supplement with boundary conditions for the fermion. If we choose $\Psi_{0,L}=\Psi_{N+1,R}=0$, the deconstructed lagrangian becomes (where we defined $e=\frac{e_4}{\sqrt{\Delta_4}}$, and dropped the primes):
\begin{equation}
\begin{aligned}
{\cal L}_{4d}= &-\frac{1}{2}\sum_{i=1}^N \overline{\Psi_i}[\gamma^\mu(\partial_\mu-ie^{\frac{ki}{2}}eA_{i,\mu})+e^{-ki}m]\Psi_i \\&-\frac{1}{2}\overline{\Psi_{0,R}}\gamma^\mu\partial_\mu\Psi_{0,R} -\frac{1}{2}\overline{\Psi_{N+1,L}}\gamma^\mu\partial_\mu\Psi_{N+1,L}\\
&-\frac{1}{2}\sum_{i=1}^{N-1}\overline{\Psi_i}\gamma_5(\frac{e^{2k}}{\Delta_4}-ie^{k(\frac{i}{2}+2)}eA_{i,4})\Psi_{i+1}\\&-\frac{1}{2}\overline{\Psi_{0,R}}(\frac{e^{2k}}{\Delta_4}-ie^{k}eA_{0,4})\Psi_{1,L}\\
&-\frac{1}{2}\overline{\Psi_{N,R}}(\frac{e^{2k}}{\Delta_4}-ie^{k(\frac{N}{2}+2)}eA_{0,4})\Psi_{N+1,L}+h.c. \ .
\end{aligned}
\end{equation}
However, we cannot UV complete this lagrangian as we did in (\ref{UVcompleted}) since its $k$-dependence prevents from recognizing the low-energy expansion of the $\phi_i$'s. Only when the background is flat ($k=0 \iff q=1$) one can follow such a procedure (when $e=1$):
\begin{equation}
\begin{aligned}
{\cal L}_{4d \text{ UV},\text{flat}}= &-\frac{1}{2}\sum_{i=1}^N \overline{\Psi_i}[\gamma^\mu(\partial_\mu+ieA_{i,\mu})+m]\Psi_i \\&-\frac{1}{2}\overline{\Psi_{0,R}}\gamma^\mu\partial_\mu\Psi_{0,R} -\frac{1}{2}\overline{\Psi_{N+1,L}}\gamma^\mu\partial_\mu\Psi_{N+1,L}\\
&-\frac{1}{\sqrt{2}}\sum_{i=1}^{N-1}\overline{\Psi_i}\gamma_5\phi_i\Psi_{i+1}-\frac{1}{\sqrt{2}}\overline{\Psi_{0,R}}\phi_0\Psi_{1,L}\\&-\frac{1}{\sqrt{2}}\overline{\Psi_{N,R}}\phi_N\Psi_{N+1,L}+h.c. \ .
\end{aligned}
\end{equation}
This lagrangian respects $U(1)_a$ but makes it anomalous at the loop level. If one now includes an allowed mass term $-\frac{m}{2}\overline{\Psi_{0,R}}\Psi_{N+1,L}+h.c.$, $U(1)_a$ is classically broken by non-local effects, which can then generate the potential $\phi_0\phi_1...\phi_N$ from fermionic loops \cite{Hill:2002me,Hill:2002kq}. When $q \neq 1$, none of this can be implemented. This reminds us that in section \ref{quarksInt} we needed $\sim q^N$ fermions to make $U(1)_a$ anomalous at the loop level, while deconstruction only provides us with $\sim N$ fermions. 

Nevertheless, in order to make $U(1)_a$ anomalous like in section \ref{quarksInt}, or to classically break it like in appendix \ref{KSVZ}, one can consider purely four dimensional setups.

\section{Massive vectors of the 4d model}\label{massive}

The model of eq. (\ref{model}) contains massive modes in addition to the Goldstone boson $a$. The vector bosons mass matrix is:
\begin{equation}
\begin{aligned}
&{\cal M}_{vect}^2=2\times\\
&\left(\begin{array}{cccc} 
q^2\abs{\phi_0}^2+\abs{\phi_1}^2& -q\abs{\phi_1}^2 & ... &0\\
-q\abs{\phi_1}^2 & q^2\abs{\phi_1}^2+\abs{\phi_2}^2  & ...&0\\
0&-q\abs{\phi_2}^2 &  ... &  -q\abs{\phi_{N-1}}^2\\
0&... & ... & q^2\abs{\phi_{N-1}}^2+\abs{\phi_N}^2
\end{array}\right)\\
&=f^2
\left(\begin{array}{cccc} 
1+q^2& -q & ... &0\\
-q & 1+q^2 & ...&0\\
0&-q & ... & -q\\
0&... & ... &1+q^2\end{array}\right)
\end{aligned} 
\end{equation}
after gauge symmetry breaking, with eigenvalues and eigenvectors:
\begin{equation}
\ba
&m_{j=1...N}^2=f^2\left(1+q^2-2q\cos(\frac{j\pi}{N+1})\right) \text{ and } \\&A'_{j=1...N}=\left(\sin(\frac{jk\pi}{N+1}),k=1...N\right) \ .
\ea
\label{clockworkmasses}
\end{equation}
All vectors are massive since all gauge symmetries are broken. We recognize in (\ref{clockworkmasses}) the specific (band-like) massive spectrum of clockwork models.

The masses of the Higgs-like $r_k$ scalar fields depend on the choices of parameters in (\ref{model}) and do not necessarily lie in a band.

\section{Realizations of benchmark QCD axion models}

We discuss the compatibility of usual benchmark invisible QCD axion models, namely KSVZ \cite{Kim:1979if,Shifman:1979if} and DFSZ \cite{Dine:1981rt,Zhitnitsky:1980tq} models, with our setup. In these models, the $U(1)_{PQ}$ anomaly with respect to QCD is respectively carried by additional heavy colored particles or by the standard model quarks, and the PQ symmetry arises from the introduction of a SM singlet scalar field (as well as an extra Higgs doublet for the DFSZ model). The phase shift symmetry of this singlet is not gauge protected in their original realization, consequently so we replace it by the accidental symmetry of our quiver model. We will also discuss, in the case of the KSVZ model, how the additional fermions can break $U(1)_a$ and generate (\ref{explicitbreaking}) as a quantum correction.

\subsection{KSVZ model: anomaly mediated by additional particles}\label{KSVZ}

The original KSVZ model was already (anonymously and briefly) introduced in (\ref{quarkInteg}), where $\sigma$ is a SM gauge singlet, and some quiver versions of it were already described in (\ref{integrateParticles}) and (\ref{higherYukawas}). There, the needed couplings were ad hoc, in contrast with the fact that we talked about an accidental Peccei-Quinn symmetry. However, we can choose the fermions charges so that the procedure of (\ref{integrateParticles}) (respectively (\ref{higherYukawas})) is automatically implied by the most general renormalizable gauge-invariant lagrangian (respectively the lowest-order gauge-invariant lagrangian which renders all the additional fermions massive),  given the gauge charges of the different fields involved. This is for instance achieved if the fermions charges are those displayed in Table \ref{colored} (respectively Table \ref{coloredreduced}).
\begin{table*}[h]
\centering
\begin{tabular}{cccccccc} 
\hline
&$U(1)_1$&$U(1)_2$&$U(1)_3$&...&$U(1)_N$&$SU(3)_c$&$U(1)_a$\\
\hline
$Q_{L,0}$&$-q$&$0$&$0$&...&$0$&$\textbf{3}$&$q_R+1$\\
\hline
$Q_{L,1}^{i=1...q}$&$1$&$-q$&$0$&...&$0$&$\textbf{3}$&$q_R+q$\\
\hline
$Q_{L,2}^{i=1...q^2}$&$0$&$1$&$-q$&...&$0$&$\textbf{3}$&$q_R+q^2$\\
\hline
...&...&...&...&...&...&...&...\\
\hline
$Q_{L,N}^{i=1...q^N}$&$0$&$0$&$0$&...&$1$&$\textbf{3}$&$q_R+q^N$\\
\hline
$Q_R^{i=1...(1+q+...q^N)}$&$0$&$0$&$0$&...&$0$&$\textbf{3}$&$q_R$\\
\hline
\end{tabular}
\caption{Colored fermions charged under the quiver gauge group of Figure \ref{quivAxion},\protect\\canceling $SU(3)_c^2-U(1)_i$ anomalies and leading to a QCD axion\protect\\($U(1)_a$ charges are those imposed by (\ref{mostgeneralKSVZ}), functions of $q_R$ which is arbitrary)}
\label{colored}
\end{table*}
\begin{table*}[h]
\centering
\begin{tabular}{ccccccc} 
\hline
&$U(1)_1$&$U(1)_2$&...&$U(1)_N$&$SU(3)_c$&$U(1)_a$\\
\hline
$Q_{L,0}$&$-q$&$0$&...&$0$&$\textbf{3}$&$q_R+1$\\
\hline
$Q_{L,1}^{i=1...q}$&$1$&$0$&...&$0$&$\textbf{3}$&$q_R+q+q^3+...+q^{q^{2N-1}}$\\
\hline
$Q_R^{i=1...(1+q)}$&$0$&$0$&...&$0$&$\textbf{3}$&$q_R$\\
\hline
\end{tabular}
\caption{Colored fermions with mass terms from higher dimensional operators\protect\\($q_R$ is arbitrary)}
\label{coloredreduced}
\end{table*}

For example, the most general renormalizable lagrangian associated with Table \ref{colored} is, with such charges:
\begin{equation}
{\cal L} \supset \ -\phi_0\overline{Q_{L,0}}Y_{0,i}Q_{R}^i-\phi_1\overline{Q_{L,1}^i}Y_{1,ij}Q_{R}^j +...+h.c. \ ,
\label{mostgeneralKSVZ}
\end{equation}
and it defines the $U(1)_a$ charges of the fermion bilinears which make $U(1)_a$ accidentally conserved, which in turn determine the $U(1)_a \times SU(3)^2$ anomaly and justify the procedure (\ref{integrateParticles}).

Along with these colored fermions, one must also add fermions only charged under the quiver gauge group to cancel the $U(1)_i \times U(1)_j \times U(1)_k$ anomalies. A way of achieving this for (\ref{integrateParticles}) is presented in Table \ref{rest}.
\begin{table*}[h]
\centering
\begin{tabular}{cccccccc} 
\hline
&$U(1)_1$&$U(1)_2$&$U(1)_3$&...&$U(1)_N$&$SU(3)_c$&$U(1)_a$\\
\hline
$\psi_{R,0}$&$-q$&$0$&$0$&...&$0$&$\textbf{1}$&$q_L+1$\\
\hline
$\psi_{R,1}^{i=1...q}$&$1$&$-q$&$0$&...&$0$&$\textbf{1}$&$q_L+q$\\
\hline
$\psi_{R,2}^{i=1...q^2}$&$0$&$1$&$-q$&...&$0$&$\textbf{1}$&$q_L+q^2$\\
\hline
...&...&...&...&...&...&...&...\\
\hline
$\psi_{R,N}^{i=1...q^N}$&$0$&$0$&$0$&...&$1$&$\textbf{1}$&$q_L+q^N$\\
\hline
$\psi_L^{i=1...(1+q+...q^N)}$&$0$&$0$&$0$&...&$0$&$\textbf{1}$&$q_L$\\
\hline
\end{tabular}
\caption{SM-singlet fermions charged under the quiver gauge group of Figure \ref{quivAxion},\protect\\canceling cubic quiver anomalies of Table \ref{colored} ($q_L$ is arbitrary)}
\label{rest}
\end{table*}

One can check at the level of these fermionic contents that the models are gauge-anomaly-free, and at the level of their most general renormalizable lagrangian that they preserve an anomalous $U(1)_a$ global symmetry.

Still, we only considered renormalizable lagrangian, so we could ask whether Planck-suppressed fermionic terms will be generated along with (\ref{explicitbreaking}), whether such terms explicitly break $U(1)_a$ and whether they can induce quantum corrections to the axion mass. In the cases discussed above, we can supplement (\ref{mostgeneralKSVZ}) by:
\begin{equation}
\ba
{\cal L} \supset \ &-\frac{\phi_1^{q*}...\phi_N^{q^N*}}{M_P^{q+...+q^N-1}}\overline{Q_{L,0}}Y'_{0,i}Q_{R}^i \\&-\frac{\phi_0^*\phi_1^{(q-1)*}\phi_2^{q^2*}...\phi_N^{q^N*}}{M_P^{q+...+q^N-1}}\overline{Q_{L,1}^{i}}Y'_{1,ij}Q_{R}^j +...+h.c. \ ,
\ea
\end{equation}
which now explicitly breaks $U(1)_a$ and induces loop corrections to $m_a^2$. However, such corrections are proportional to the factor $\frac{Y'}{M_P^{q+...+q^N-1}}$ since $U(1)_a$ is perturbatively preserved when those terms are equal to zero. Hence, by comparing with (\ref{mass}) where $m_a^2 \sim \frac{1}{M_P^{q+...+q^N-3}}$, we conclude that (\ref{mass}) gives the leading contribution to the axion mass.\footnote{When one takes into account the $\psi$ fields of Table \ref{rest}, one could also write gauge-invariant Majorana mass terms for the $\psi_L$'s, but these do not break $U(1)_a$.}

However, this conclusion depends on the choice of gauge charges. For instance, if one chooses $q=3,N=2$ and the gauge charges of Table \ref{coloredBreaking}, one can write the following lagrangian:
\begin{table}[h]
\centering
\begin{tabular}{ccccccc} 
  \hline
&$U(1)_1$&$U(1)_2$&$SU(3)_c$&$U(1)_a$\\
\hline
$Q_{L,0}$&$0$&$0$&$\textbf{3}$&$q_0$\\
\hline
$Q_{R,0}$&$3$&$0$&$\textbf{3}$&$q_0-1$\\
\hline
$Q_{L,1}^{i=1...3}$&$1$&$0$&$\textbf{3}$&$q_0+q+q^3$\\
\hline
$Q_{R,1}^{i=1...3}$&$0$&$3$&$\textbf{3}$&$q_0+q^3$\\
\hline
$Q_{L,2}^{i=1...9}$&$0$&$1$&$\textbf{3}$&$q_0+q^2$\\
\hline
$Q_{R,2}^{i=1...9}$&$0$&$0$&$\textbf{3}$&$q_0$\\
\hline
\end{tabular}
\caption{Colored fermions giving the major contribution to the axion mass\protect\\($U(1)_a$ charges are those imposed by (\ref{lagColoredBreaking}), functions of $q_0$ which is arbitrary))}
\label{coloredBreaking}
\end{table}
\begin{equation}
\begin{aligned}
{\cal L} \supset& -\overline{Q_{L,0}}MQ_{R,2}-\phi_0\overline{Q_{L,0}}Y_{00}Q_{R,0}-\phi_1\overline{Q_{L,1}}Y_{11}Q_{R,1}\\
&-\phi_2\overline{Q_{L,2}}Y_{22}Q_{R,2}-\frac{\phi_2^{2*}}{M_P}\overline{Q_{L,2}}Y_{21}Q_{R,1}\\
&-\frac{\phi_1^{2*}\phi_2^{6*}}{M_P^7}\overline{Q_{L,1}}Y'_{10}Q_{R,0}+h.c.
\label{lagColoredBreaking}
\end{aligned}
\end{equation}
(where we omitted flavour indices and some gauge invariant terms which do not break $U(1)_a$ and thus have no impact on the discussion). The five first terms of (\ref{lagColoredBreaking}) fix the $U(1)_a$ charges displayed in Table \ref{coloredBreaking}, whereas the last one breaks this charge assignment since it has a global charge $-1-q^2-q^4$. However, as soon as one of the $M,Y^{(')}$ is zero, $U(1)_a$ is conserved. Consequently, (\ref{lagColoredBreaking}) induces a loop correction to $m_a^2$ proportionnal to $\frac{1}{M_P^8}$ whereas the square of (\ref{mass}) is proportionnal to $\frac{1}{M_P^9}$. Thus, in this case, gravitational corrections to the fermion lagrangian induce a mass for the axion which competes with the pure scalar breaking of (\ref{explicitbreaking}).

\subsection{DFSZ model: anomaly mediated by standard model quarks}\label{DFSZ}

We focus now on the DFSZ model, since, contrary to the KSVZ model, the original model has the important feature that the anomaly is only carried by the standard model quarks. It makes uses of two Higgs doublets $H_{1,2}$, an extra singlet scalar $\sigma$ and can be summarized as follows:
\begin{equation}
\begin{aligned}
{\cal L} \supset& -\overline{u_R}H_1Y_uQ_L - \overline{d_R}H_2Y_dQ_L \\
&\hphantom{+++}- \overline{e_R}H_2Y_eL_L-\lambda H_1H_2\sigma^2\\ 
&\xrightarrow[]{u,d \text{ triangles}}  \frac{i}{32\pi^2}\log(H_1H_2)G\tilde{G} -\lambda H_1H_2\sigma^2\ .
\label{DFSZsummary}
\end{aligned}
\end{equation}
The first line of (\ref{DFSZsummary}) is invariant under a global $U(1)$ which acts on the scalars as $\sigma \rightarrow e^{i\alpha}\sigma,H_{1,2}\rightarrow e^{-i\alpha}H_{1,2}$. The symmetry is spontaneously broken and, according to the second line of (\ref{DFSZsummary}), anomalous with respect to QCD.

In order to adapt this construction to the case of our quiver, it is important to disentangle two features of (\ref{DFSZsummary}): the $\log(H_1H_2)$ operator originates from the Yukawa terms of the SM quarks which run into loops,\footnote{Actually, since three quark families run in the loops, the correct operator is $\log((H_1H_2)^3)$.} whereas the $H_1H_2\sigma^2$ term (and the rest of the tree-level lagrangian) defines which symmetry is respected.\footnote{and thus which combination of the phases of the scalars is a genuine massless Goldstone boson. If $\sigma$ is assumed to get an intermediate scale vev, this boson is mostly located on the phase of $\sigma$ and evades the astrophysical constraints on an electroweak scale axion.} Thus, if we want to apply this logic to $U(1)_a$, we must identify gauge charges of $H_1$ and $H_2$ which will preserve the accidental $U(1)_a$, and identify a gauge-invariant operator $O$, charged under $U(1)_a$, which will induce an axionic coupling $\log(O)G\tilde{G}$ to the gluons. We can immediately understand from section \ref{quarksInt} that $O$ must be of high dimension, so it must be generated by more colored particles than standard model quarks alone. It would thus be more precise to talk about a mixed DFSZ-KSVZ model, where the anomaly is mediated by both standard model quarks and additional fermions. In particular, we loose the pleasant economical quark content of the original DFSZ model, since one needs a growing number of additional particles as in the KSVZ case.

As an (unoptimized) example of this procedure, we choose the matter content and gauge charges of Table \ref{DFSZmodel} in addition to that of Figure \ref{quivAxion}.\footnote{All anomalies involving a standard model factor are canceled. The cubic, as well as the mixed abelian-gravitational anomalies of the quiver gauge group can be canceled by adding heavy SM-singlet fermions with charges identical to those of the additional fermions in Table \ref{DFSZmodel}, with SM representations turned into multiplicities, in the spirit of Table \ref{rest}.}
\begin{table*}[h]
\centering
\begin{tabular}{ccccccccc}
\hline
$\text{Fields}$&$SU(3)_c$&$SU(2)_{EW}$&$U(1)_Y$&$U(1)_1$&$U(1)_2$&$U(1)_3$&...&$U(1)_a$\\
\hline
$Q_L$&$\textbf{3}$&$\textbf{2}$&$\dfrac{\strut 1}{\strut 6}$&$-\dfrac{\strut 4q}{\strut3}$&$0$&$0$&...&$q_Q$\\
\hline
$u_R$&$\textbf{3}$&$\textbf{1}$&$\dfrac{\strut2}{\strut3}$&$-\dfrac{\strut q}{\strut3}$&$0$&$0$&...&$q_Q+q_H$\\
\hline
$d_R$&$\textbf{3}$&$\textbf{1}$&$-\dfrac{\strut1}{\strut3}$&$-\dfrac{\strut q}{\strut3}$&$0$&$0$&...&$q_Q-q_H-2$\\
\hline
$L_L$&$\textbf{1}$&$\textbf{2}$&$-\dfrac{\strut1}{\strut2}$&$-\dfrac{\strut 4q}{\strut3}$&$0$&$0$&...&$q_e+q_H+2$\\
\hline
$e_R$&$\textbf{1}$&$\textbf{1}$&$-1$&$-\dfrac{\strut q}{\strut3}$&$0$&$0$&...&$q_e$\\
\hline
$H_1$&$\textbf{1}$&$\textbf{2}$&$\dfrac{\strut1}{\strut2}$&$q$&$0$&$0$&...&$q_H$\\
\hline
$H_2$&$\textbf{1}$&$\textbf{2}$&$-\dfrac{\strut1}{\strut2}$&$q$&$0$&$0$&...&$-q_H-2$\\
\hline
$Q_{L,\text{EW}}^{i=1...16}$&$\textbf{1}$&$\textbf{2}$&$0$&$q$&$0$&$0$&...&$q_{EW}$\\
\hline
$Q_{R,\text{EW}}^{i=1...16}$&$\textbf{1}$&$\textbf{2}$&$0$&$0$&$0$&$0$&...&$q_{EW}+1$\\
\hline
$Q_{R,0}^{i=1...5}$&$\textbf{3}$&$\textbf{1}$&$0$&$-q$&$0$&$0$&...&$q_L-1$\\
\hline
$Q_{R,1}^{i=1...q}$&$\textbf{3}$&$\textbf{1}$&$0$&$1$&$-q$&$0$&...&$q_L-q$\\
\hline
$Q_{R,2}^{i=1...q^2}$&$\textbf{3}$&$\textbf{1}$&$0$&$0$&$1$&$-q$&...&$q_L-q^2$\\
\hline
...&...&...&...&...&...&...\\
\hline
$Q_{L}^{i=1...(5+q+q^2+...q^N)}$&$\textbf{3}$&$\textbf{1}$&$0$&$0$&$0$&$0$&...&$q_L$\\
\hline
\end{tabular}
\caption{Matter content for the quiver DFSZ model\protect\\($U(1)_a$ charges are those imposed by (\ref{DFSZlag}), functions of $q_Q,q_H,q_e,q_{EW}$ and $q_L$ which are arbitrary)}
\label{DFSZmodel}
\end{table*}
With these charges, one has the following most general renormalizable interaction terms:
\begin{equation}
\begin{aligned}
&{\cal L} \supset -\overline{u_R}H_1Y_uQ_L -\overline{d_R}H_2Y_dQ_L -\overline{e_R}H_2Y_eL_L\\
&\hphantom{{\cal L} \supset}-\phi_0\overline{Q_{R,\text{EW}}^{i}}Y_{\text{EW},ij}Q_{L,\text{EW}}^j -\phi_0\overline{Q_{R,0}^{i}}Y_{0,ij}Q_{L}^j\\
&\hphantom{{\cal L} \supset}-\phi_1^* \overline{Q_{R,1}^{i}}Y_{1,ij}Q_{L}^j-\phi_2^* \overline{Q_{R,2}^{i}}Y_{2,ij}Q_{L}^j-...\\
&\hphantom{{\cal L} \supset}-\lambda H_1H_2\phi_0^2+h.c.\\
&\xrightarrow[]{\text{ triangles}}  -\frac{i}{32\pi^2}\log((H_1H_2)^3\phi_0^5\phi_1^{*q}\phi_2^{*q^2}...\phi_N^{*q^N})G\tilde{G}\\&\hphantom{\xrightarrow[]{\text{ triangles}},}-(\lambda H_1H_2\phi_0^2+h.c.) \ ,
\end{aligned}
\label{DFSZlag}
\end{equation}
where we identify $O=(H_1H_2)^3\phi_0^5\phi_1^{*q}\phi_2^{*q^2}...\phi_N^{*q^N}$. $U(1)_a$ charges are assigned to $H_{1,2}$ so that $H_1H_2\phi_0^2$ is invariant, and $\log(O)G\tilde{G}$ makes $U(1)_a$ anomalous. The axion effective decay constant displays the same asymptotic dependence than (\ref{fa}): $f_a\sim \frac{f}{q^N}$.

It is worth noticing that a $\mu^2 H_1H_2$ or $\mu H_1H_2\sigma$ term was not included in (\ref{DFSZsummary}) in order to maintain a global symmetry, whereas we now cannot write something else than (\ref{DFSZlag}) that would respect gauge symmetries, which was the original goal when we introduced the quiver. The first allowed $U(1)_a$-violating operator is again $\phi_0\phi_1^q...\phi_N^{q^N}$ and the discussion around eq. (\ref{protectQCD}) applies.

\section{Couplings of the axion to gauge vectors}\label{calculation}

We compute the axion-photon-photon coupling for the model of Figure \ref{quivAxion} and Table \ref{ALPtogamma}. However, the calculation performed here is very general and can also be seen as a derivation of (\ref{quarkInteg}). 

One considers first a theory with a gauge group (which we keep unspecified until the end, where we will identify it with QCD or electromagnetism) of generators $T^a$, coupling constant $g$ and vector $A_\mu^a$ (with field strength $F_{\mu\nu}^a=\partial_\mu A_\nu^a-\partial_\nu A_\mu^a+...$), a complex scalar field $\sigma$ and two chiral fermions $\psi_{L,R}$ in the fundamental representation of the gauge group, with a Yukawa coupling to the scalar:
\begin{equation}
\ba
{\cal L} =& -\frac{1}{4g^2}F_{\mu\nu,a}^2-\overline{\psi_L}\gamma^\mu D_\mu \psi_L-\overline{\psi_R}\gamma^\mu D_\mu \psi_R\\
&-\abs{\partial\sigma}^2-V(\abs{\sigma}^2) -(y\sigma\overline{\psi_L}\psi_R+h.c.) \ ,
\ea
\label{appendixLag}
\end{equation}
where $D_\mu=\partial_\mu-iA_\mu^aT^a$. This lagrangian has a $U(1)$ global symmetry under which $\sigma \rightarrow e^{i\alpha}\sigma$ and $\overline{\psi_L}\psi_R \rightarrow e^{-i\alpha}\overline{\psi_L}\psi_R$. The transformation of the fermion bilinear makes this global symmetry anomalous. 

We choose $V(\abs{\sigma}^2)$ so that $\sigma$ gets a vev $f$. We then work out the axion dynamics by parametrizing $\sigma=\frac{f}{\sqrt{2}}e^{i\frac{a}{f}}$:
\begin{equation}
{\cal L} \supset -\frac{1}{4g^2}F_{\mu\nu}^2-\overline{\psi}\gamma^\mu (D_\mu+\frac{yf}{\sqrt{2}}) \psi-\frac{1}{2}(\partial a)^2 +i\frac{y}{\sqrt{2}}a\overline{\psi}\gamma_5\psi \ ,
\end{equation}
where we only kept the linear terms in $a$ and merged the two chiral fermions in a Dirac fermion.

\begin{figure*}[h]
\centering
\includegraphics[scale=0.4]{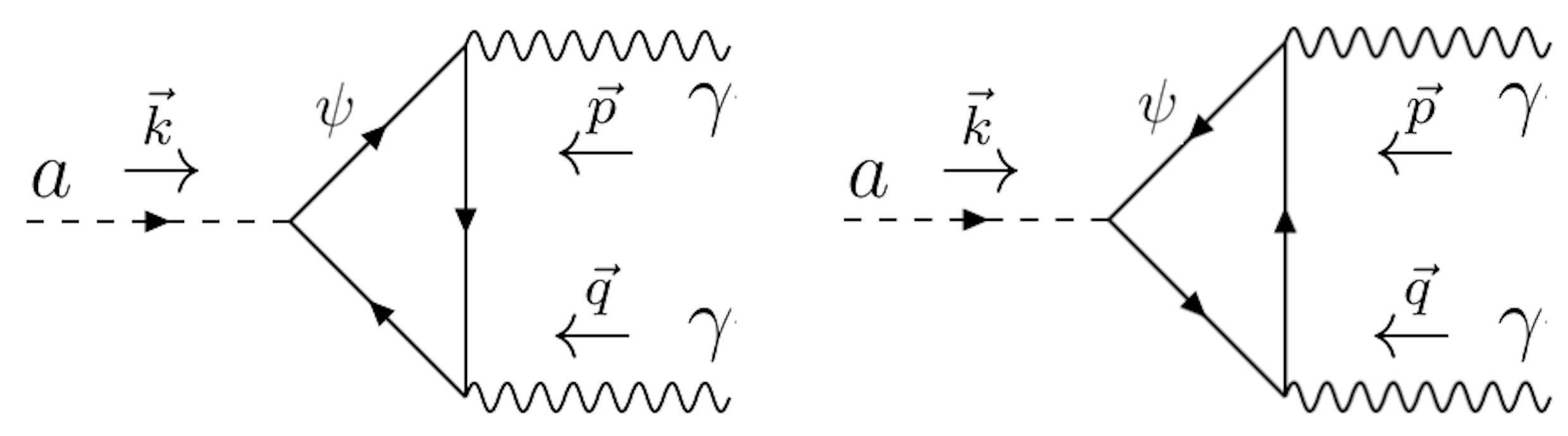}
\caption{Feynman diagrams leading to the axion-vector-vector couplings}
\label{triangleAxion}
\end{figure*}

One gets a coupling between the axion $a$ and the gauge boson $A$ at one loop via the diagrams of Figure \ref{triangleAxion}.

The effective coupling is $c^{\mu\nu,ab}aA_\mu^aA_\nu^b$, here in momentum space with $M_\psi=\frac{yf}{\sqrt{2}}$ and at first order in $\frac{p}{M_\psi},\frac{q}{M_\psi}$:
\begin{equation}
c^{\mu\nu,ab}= \frac{-1}{4\pi^2f}\delta^{ab}\epsilon^{\mu\nu\rho\sigma}p_\rho q_\sigma \Big(\frac{1}{2}-\frac{p^2+q^2+pq}{12M_\psi^2}\Big)\\
\end{equation}
which, with the identification $pA(p) \rightarrow -i\partial A(x)$, gives finally the one-loop coupling between the axion and the vector bosons:
\begin{equation}
\ba
{\cal L} \supset &-\frac{\epsilon^{\mu\nu\rho\sigma}}{32\pi^2f}aF_{\mu\nu}^aF_{\rho\sigma}^a\\
&+\frac{\epsilon^{\mu\nu\rho\sigma}}{192\pi^2M_\psi^2f}(-\Box aF_{\mu\nu}^aF_{\rho\sigma}^a+2\partial_\mu a \partial^\eta F_{\rho\sigma}^aF_{\nu\eta}^a) \ .
\ea
\label{appendixResult}
\end{equation}
The first term of (\ref{appendixResult}) is the usual axionic coupling to gauge fields, while the other terms match similar calculations already performed in the literature (see for example \cite{Dudas:2013sia}).

If one now adds to the theory (\ref{appendixLag}) another set of fermions coupled in the following way:
\begin{equation}
\begin{aligned}
{\cal L} \supset& -\overline{\psi'_L}\gamma^\mu D_\mu \psi'_L-\overline{\psi'_R}\gamma^\mu D_\mu \psi'_R-(y'\sigma^*\overline{\psi'_L}\psi'_R+h.c.)\\
&\xrightarrow[]{\text{axion terms}} -\overline{\psi'}\gamma^\mu (D_\mu+\frac{y'f}{\sqrt{2}}) \psi'-i\frac{y'}{\sqrt{2}}a\overline{\psi'}\gamma_5\psi' \ ,
\end{aligned}
\end{equation}
there is no anomaly anymore, but there remains non-anomalous couplings to the gauge fields (where we defined $M_\psi'=y'f$):
\begin{equation}
{\cal O} \supset \frac{\epsilon^{\mu\nu\rho\sigma}}{192\pi^2f}(\frac{1}{M_\psi^2}-\frac{1}{M_\psi'^2})(-\Box aF_{\mu\nu}^aF_{\rho\sigma}^a+2\partial_\mu a \partial^\eta F_{\rho\sigma}^aF_{\nu\eta}^a) \ .
\end{equation}
Specializing to electromagnetism, normalizing the photon field $A_\mu \rightarrow e A_\mu$ and choosing $\sigma=\phi_i=\frac{f}{\sqrt{2}}e^{i\frac{q^ia/f}{\sqrt{1+...+q^{2N}}}}$, one obtains (\ref{atogamma}).

\section{Scan of the parameters which allow for (detectable) ALP DM}\label{scanAll}

\begin{figure*}[h]
\centering
\includegraphics[scale=0.5]{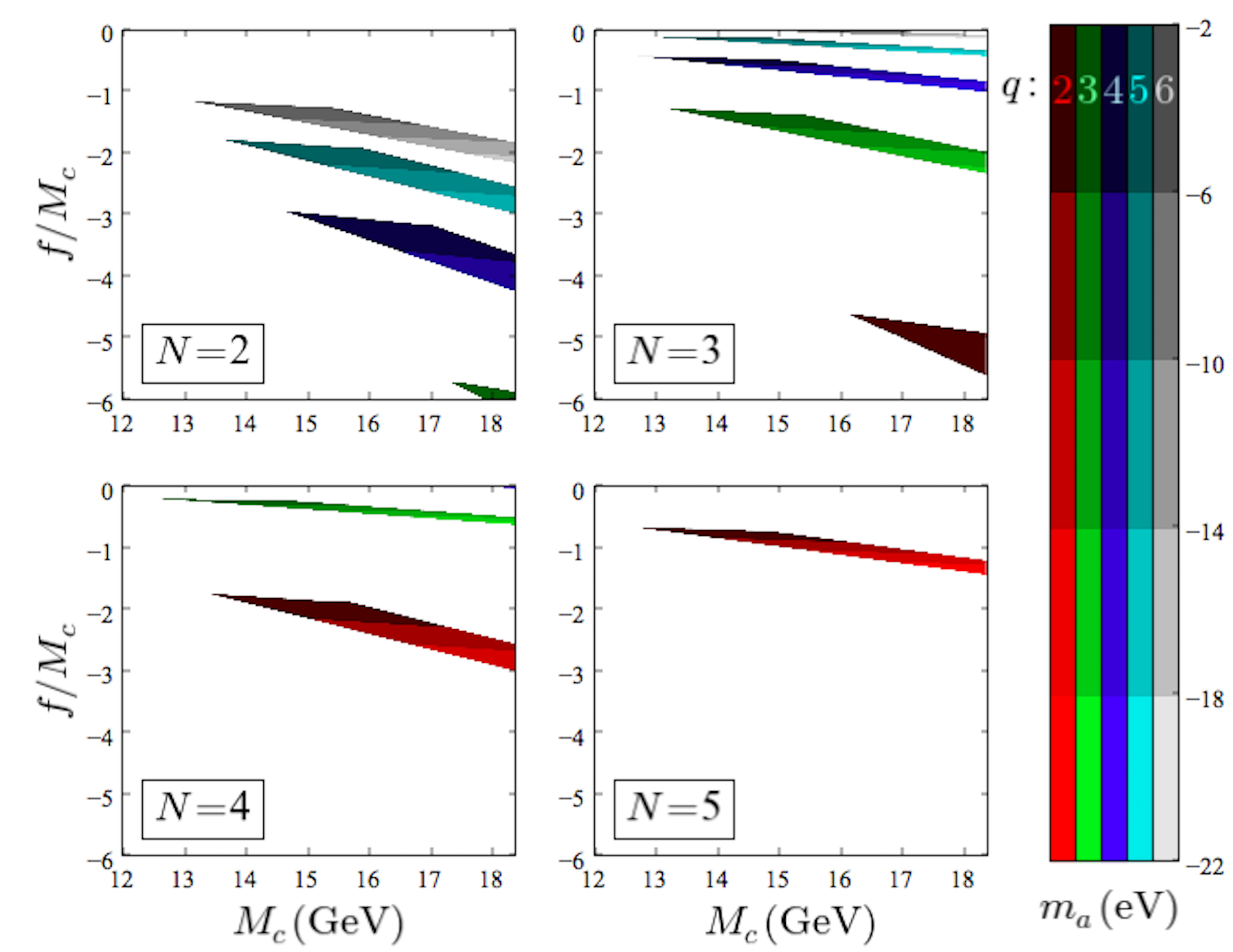}
\caption{Range of parameters for a DM ALP of mass $m_a\leq 10^{-2}$ eV\protect\\(axions suitable to saturate the DM relic density are found in colored regions, all axes are log-scale)}
\label{moreDM}
\end{figure*}

We extend in this appendix the analysis performed in section \ref{ALPs} to more values of $q$ and $N$, since the DM examples in Figure \ref{scanALPs} have been arbitrarily chosen. Figure \ref{moreDM} displays all DM candidates in our setup when $q \leq 6$ and $N \leq 5$, with tuning restrictions identical to those used in Figure \ref{scanALPs}. As mentioned in section \ref{ALPs}, those results were obtained assuming that $U(1)_a$ was broken above the inflation scale. Indeed, we can see from Figures \ref{scanALPs}, \ref{zoomMP} and \ref{moreDM} that most of our ALP DM candidates require $f$ to be high (whereas the inflation scale, given by the Hubble rate during inflation, verifies $H_{\text{inflation}} \lesssim 10^{14} \text{ GeV}$). Consequently, we only focus on the broken case (which may suffer from isocurvature fluctuations issues, which are however negligible when $f$ is close to $M_P$).

\begin{figure*}[h]
\centering
\begin{tabular}{c} 
\includegraphics[scale=0.5]{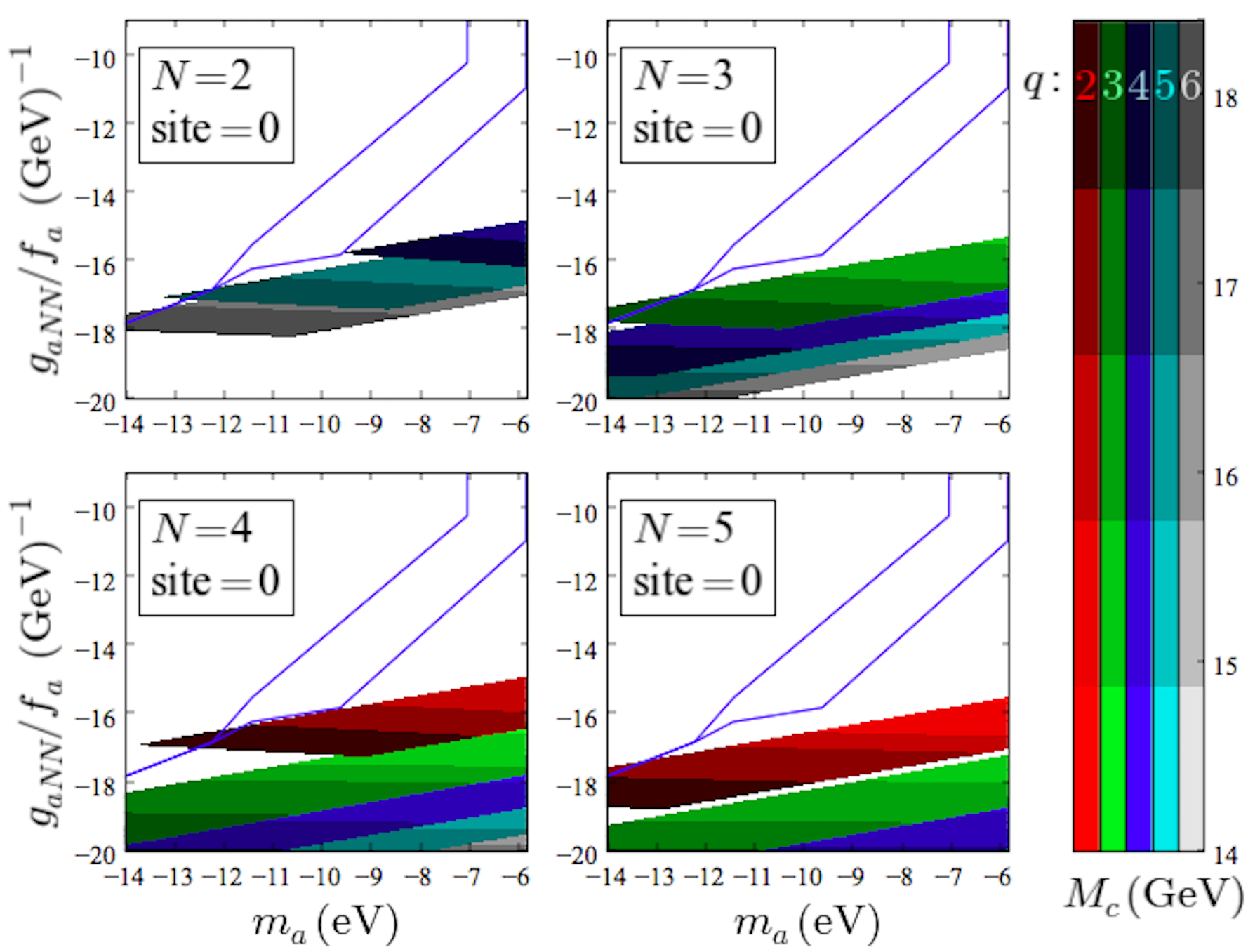}
\\
\includegraphics[scale=0.5]{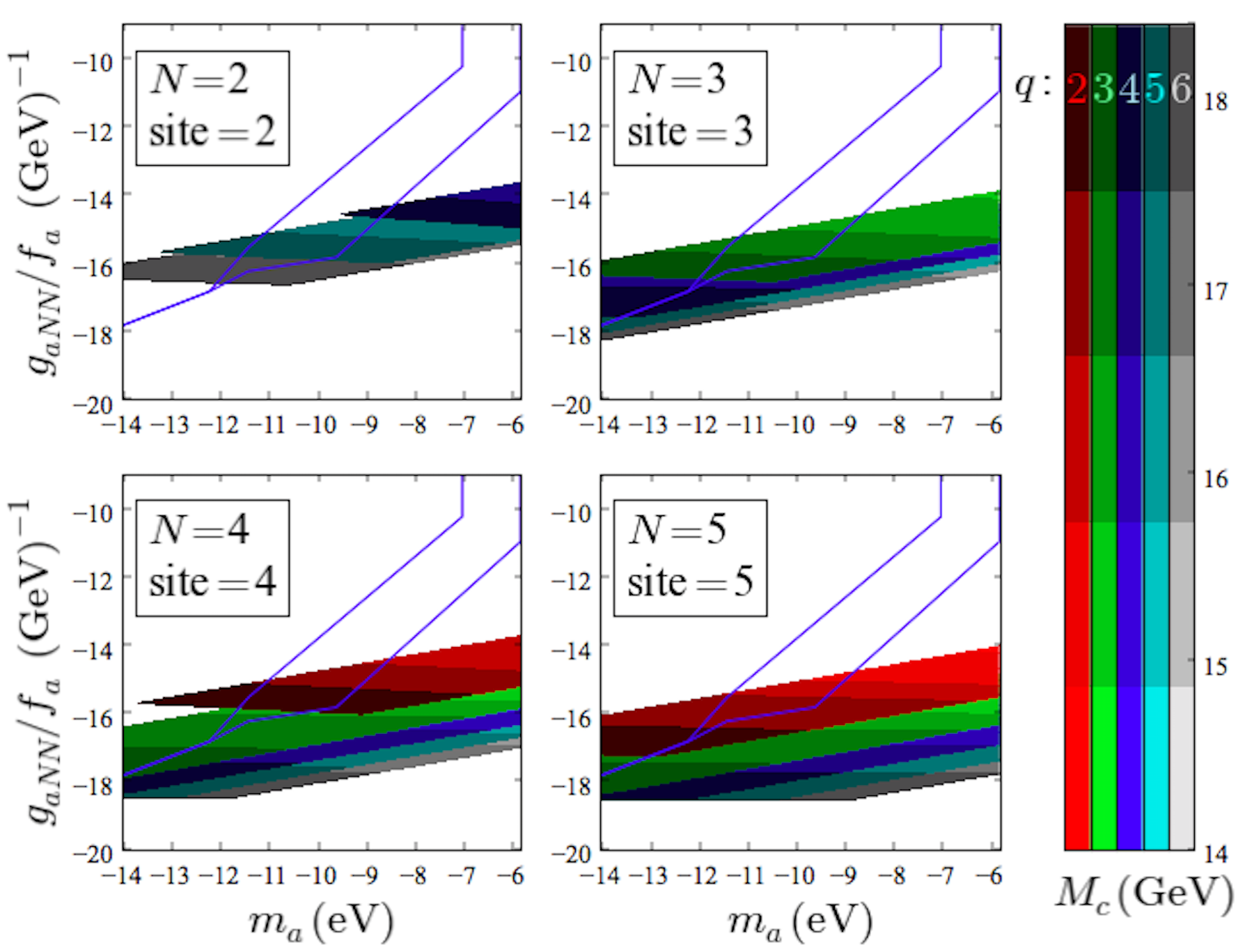}
\end{tabular}
\caption{Sensitivity of CASPEr-Wind to the ALPs\protect\\(colored regions indicate axions suitable to saturate the DM relic density, detection happens in the upper left part of the plot, blue lines are identical to those of Figure \ref{CasperSensitivity}, all axes are log-scale)}
\label{CasperSensitivityMore}
\end{figure*}

We also allow in Figure \ref{CasperSensitivityMore} (which, as Figure \ref{CasperSensitivity}, compares the sensitivity of the CASPEr-Wind experiment with the predictions of our model) for more values of $q$ and $N$, but also for $M_c<M_P$. The upper panel of Figure \ref{CasperSensitivityMore} couples the standard model with the first site of the quiver while the lower panel couples it to the last site of the quiver (which, as visible in the plot, increases the coupling and thus the detectability of the setup). We see from Figure \ref{CasperSensitivityMore} that CASPEr-Wind experiments are more sensitive to high scale (e.g. gravitational) values of $M_c$.

\end{appendices}

\bibliographystyle{spphys}
\bibliography{AxionsInAHighlyProtectedGaugeSymmetryModel.bib}

\end{document}